\documentclass[conference,compsoc]{IEEEtran}
\IEEEoverridecommandlockouts
\usepackage{cite}
\usepackage{amsmath,amssymb,amsfonts}
\usepackage{algorithmic}
\usepackage{graphicx}
\usepackage{textcomp}
\usepackage[table]{xcolor}
\usepackage{tabularx}
\usepackage{booktabs}
\usepackage{multirow}
\usepackage{multicol}
\usepackage{comment}
\usepackage{fdsymbol}
\usepackage{url}

\usepackage[T1]{fontenc}

\usepackage{enumitem}

\usepackage{hyperref}
\usepackage{cleveref}
\Crefformat{figure}{#2Fig.~#1#3}
\Crefmultiformat{figure}{Figs.~#2#1#3}{ and~#2#1#3}{, #2#1#3}{ and~#2#1#3}
\Crefformat{section}{#2Sec.~#1#3}
\Crefmultiformat{section}{Secs.~#2#1#3}{ and~#2#1#3}{, #2#1#3}{ and~#2#1#3}
\Crefformat{table}{#2Tab.~#1#3}
\Crefmultiformat{table}{Tabs.~#2#1#3}{ and~#2#1#3}{, #2#1#3}{ and~#2#1#3}
\hypersetup{
  colorlinks,
  linkcolor =					blue,
  citecolor =					blue,
  urlcolor =					blue,
  plainpages =        false,
  hypertexnames =     true,
  linktocpage =       false,
  bookmarksopen =     true,
  bookmarksnumbered = true,
  bookmarksopenlevel= 0
}

\def\BibTeX{{\rm B\kern-.05em{\sc i\kern-.025em b}\kern-.08em
    T\kern-.1667em\lower.7ex\hbox{E}\kern-.125emX}}

\makeatletter
\newcommand{\linebreakand}{%
\end{@IEEEauthorhalign}
\hfill\mbox{}\par
\mbox{}\hfill\begin{@IEEEauthorhalign}
}
\makeatother

\newif\ifrev
\revfalse
\ifrev 
\newcommand{\zb}[1]{{~}{\color{blue} [Zina: #1]}}
\newcommand{\jw}[1]{{~}{\color{red} [Julia: #1]}}
\newcommand{\ce}[1]{{~}{\color{magenta} [Christian: #1]}}
\newcommand{\fg}[1]{{~}{\color{violet} [Freya: #1]}}
\newcommand{\ak}[1]{{~}{\color{darkspringgreen} [Andreas: #1]}}
\else 
\newcommand{\zb}[1]{}
\newcommand{\jw}[1]{}
\newcommand{\ce}[1]{}
\newcommand{\fg}[1]{}
\newcommand{\ak}[1]{}
\fi

\definecolor{darkspringgreen}{rgb}{0.00, 0.40, 0.00}

\begin{document}

\title{Shedding Light on CVSS Scoring Inconsistencies: A User-Centric Study on Evaluating Widespread Security Vulnerabilities}
\IEEEspecialpapernotice{\footnotesize Authors' version; to appear in the Proceedings of the IEEE Symposium on Security and Privacy (S\&P) 2024}
\IEEEaftertitletext{\vspace{-1\baselineskip}}

\author{
    \IEEEauthorblockN{
    Julia Wunder\IEEEauthorrefmark{1},
    Andreas Kurtz\IEEEauthorrefmark{2},
    Christian Eichenmüller\IEEEauthorrefmark{1},
    Freya Gassmann\IEEEauthorrefmark{3}
    and Zinaida Benenson\IEEEauthorrefmark{1}
    }
    \IEEEauthorblockA{
    \IEEEauthorrefmark{1}Friedrich-Alexander-Universität Erlangen-Nürnberg, firstname.lastname@fau.de
    }
    \IEEEauthorblockA{
    \IEEEauthorrefmark{2}Heilbronn University of Applied Sciences, andreas.kurtz@hs-heilbronn.de
    }
    \IEEEauthorblockA{
    \IEEEauthorrefmark{3}Rheinland-Pfälzische Technische Universität Kaiserslautern-Landau, freya.gassmann@rptu.de
    }
}

\maketitle

\begin{abstract}
The Common Vulnerability Scoring System (CVSS) is a popular method for evaluating the severity of vulnerabilities in vulnerability management. In the evaluation process, a numeric score between 0 and 10 is calculated, 10 being the most severe (critical) value. The goal of CVSS is to provide comparable scores across different evaluators. However, previous works indicate that CVSS might not reach this goal: If a vulnerability is evaluated by several analysts, their scores often differ. This raises the following questions: Are CVSS evaluations consistent? Which factors influence CVSS assessments? We systematically investigate these questions in an online survey with 196 CVSS users. We show that specific CVSS metrics are inconsistently evaluated for widespread vulnerability types, including Top 3 vulnerabilities from the ``2022 CWE Top 25 Most Dangerous Software Weaknesses'' list. In a follow-up survey with 59 participants, we found that for the same vulnerabilities from the main study, 68\% of these users gave different severity ratings. Our study reveals that most evaluators are aware of the problematic aspects of CVSS, but they still see CVSS as a useful tool for vulnerability assessment. Finally, we discuss possible reasons for inconsistent evaluations and provide recommendations on improving the consistency~of~scoring.
\end{abstract}

\section{Introduction}
The Common Vulnerability Scoring System (CVSS) is a popular system for evaluating the severity of vulnerabilities. In the vulnerability assessment process, the metrics of CVSS, which reflect different characteristics of vulnerabilities, are evaluated by human users, and then a numeric score -- the CVSS Base Score -- is computed from the values of the metrics using a specified formula. CVSS is used by many companies and organizations as an industry standard~\cite{2011-wang,2009-scarfone}. For example, the well-known vulnerability database NVD\footnote{\url{https://nvd.nist.gov}} provides CVSS scores for each vulnerability. 

Scoring the severity of vulnerabilities is a crucial step in vulnerability management, as it helps to decide whether and how promptly countermeasures must be taken to prevent exploitation. If the severity is overestimated, resources may be misallocated to remediate it. On the other hand, a vulnerability with underestimated severity may cause damage, as it may remain unpatched for too long. 
CVSS thus bears a great responsibility, and consistent severity evaluations are a cornerstone for efficiently dealing with vulnerabilities.

However, previous research has shown that scores of different evaluators are likely to differ~\cite{2020-allodi,2015-holm}, but little is still known about the factors that influence the scoring. 
CVSSv3.1 Specification~\cite[p. 5]{cvssspecsv3.1} states: ``[...] it is likely that many different types of individuals will be scoring vulnerabilities (e.g., software vendors, vulnerability bulletin analysts, security product vendors), however, note that vulnerability scoring is intended to be agnostic to the individual and their organization.''

\begin{figure*}
	\begin{center}
		\includegraphics[width=0.95\textwidth]{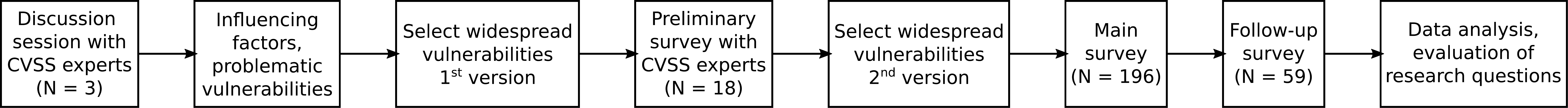}
	\end{center}
	\caption{Overview of study design ($N$ = number of participants).}
	\label{fig:outline}
\end{figure*}

We conducted a systematic investigation of CVSSv3.1 (the most current version of CVSS) as depicted in \Cref{fig:outline}.

Through analysis of related work and a discussion session with three experienced CVSS evaluators, we identified possible factors influencing CVSS assessments, problematic parts of CVSS documentation and widespread vulnerability types where CVSS evaluations might be inconsistent.
Next, we defined research questions and systematically selected common and impactful candidate vulnerabilities to be evaluated. In a preliminary online survey with 18 experienced CVSS evaluators we gathered additional expert opinion and tested the suitability of candidate vulnerabilities. In the main study, we conducted an online survey with 196 CVSS users. 
We also gathered free-text comments on possible issues with the evaluated vulnerabilities, and on CVSS usage and attitudes.
Finally,  we conducted a follow-up online survey with 59 participants, who evaluated the same vulnerabilities that they evaluated in the main study to investigate the consistency of evaluations over time.
To the best of our knowledge, this is the first study that systematically evaluates the consistency of CVSS ratings for common and impactful vulnerability types.

The contributions of this study are as follows:

\begin{enumerate}
	\item We show that CVSS metrics Attack Vector, User Interaction and Scope were not consistently assessed by 196 professional CVSS users for widespread vulnerabilities such as Out-of-bounds Write (2022 CWE Top 1)\footnote{\url{https://cwe.mitre.org/top25/archive/2022/2022_cwe_top25.html}} and Cross-site Scripting (2022 CWE Top 2).
	\item We provide evidence that factors influencing evaluation correctness are more closely related to the properties of CVSS, such as problematic metrics, than to personal characteristics of evaluators such as work experience or knowledge of CVSS documents.
	\item We show that security deficiencies (e.g., Missing HTTPOnly Flag, Banner Disclosure) were also inconsistently evaluated and led to comments on their suitability for being evaluated with CVSS.
	\item Moreover, in the follow-up study with 59 participants (volunteers from the main study) conducted  9 months later, evaluations of the same person changed severity for 68\% of participants.
	\item We shed light on the CVSS scoring process, e.g., on average 10 evaluators are usually involved per evaluation, CVSS assessments take about 5 minutes on average, and the online calculator is the most often used tool.
	\item We show that CVSS documentation is consulted rarely, and 30\% of users in our sample have never read it.
	\item We analyze attitudes towards CVSS, e.g., although 85\% of evaluators in our sample find CVSS inconsistent, most participants (80\%) still find it a useful tool for vulnerability assessment.
	\item We release a public data set with questionnaires, descriptive results and pseudonymized datasets from the main study and follow-up study \cite{public-data-set}.
\end{enumerate}

\emph{Outline.} 
This paper is organized as follows. \Cref{sec:background} gives an overview of CVSS and related work. \Cref{sec:development-research-questions} outlines an expert discussion and an expert survey we conducted to understand problematic aspects of CVSS. Research questions are  presented in \Cref{sec:research-questions}.
In \Cref{sec:methods} design of the main survey and analysis methods are specified, followed by  presentation of the results in \Cref{sec:results}.
We discuss our results in \Cref{sec:discussion},  and conclude in \Cref{sec:conclusion}.
\section{Background and Related Work}
\label{sec:background}
\begin{table*}[h!]
	\begin{tabularx}{\linewidth}{p{3cm}p{10cm}p{4cm}}
		\toprule
		Metric & Description in FIRST Specification Document & Values\\
		\midrule
		Attack Vector (AV) & ''[...] reflects the context by which vulnerability exploitation is possible.`` (p. 5) & \mbox{Network (N), Adjacent (A),} \mbox{Local (L),} Physical (P)\\
		Attack Complexity (AC) & ''[...] describes the conditions beyond the attacker's control that must exist in order to exploit the vulnerability.`` (p. 6) & Low (L), High (H)\\
		Privileges Required (PR) & ''[...] describes the level of privileges an attacker must possess before successfully exploiting the vulnerability.`` (p. 7) & None (N), Low (L), High (H)\\
		User Interaction (UI) & ''[...] captures the requirement for a human user, other than the attacker, to participate in the successful compromise of the vulnerable component.`` (p. 8) & None (N), Required (R)\\
		\midrule
		Scope (S) & ''[...] captures whether a vulnerability in one vulnerable component impacts resources in components beyond its security scope.`` (p. 8) & Unchanged (U), Changed (C)\\
		\midrule
		Confidentiality (C) & ''[...] measures the impact to the confidentiality of the information resources managed by a software component due to a successfully exploited vulnerability.`` (p. 10) & High (H), Low (L), None (N)\\
		Integrity (I) & ''[...] measures the impact to integrity of a successfully exploited vulnerability.`` (p. 10) & High (H), Low (L), None (N)\\
		Availability (A) & ''[...] measures the impact to the availability of the impacted component resulting from a successfully exploited vulnerability.`` (p. 10) & High (H), Low (L), None (N)\\
		\bottomrule
	\end{tabularx}
	\vspace*{0.5em}
	\caption{CVSS metrics and their values~\cite{cvssspecsv3.1}.}
	\label{tab:CVSSmetrics}
\end{table*} 

\subsection{CVSS Specification and Usage}
\label{sec:cvss-specification}
 
CVSS was developed by the National Infrastructure Advisory Council (NIAC)\footnote{\url{https://www.cisa.gov/niac}} and is currently maintained by FIRST\footnote{\url{https://www.first.org}}~\cite{cvss_faq}.
CVSS consists of three scores which can take any numeric value between 0.0 and 10.0: The Base Score, the Environmental Score and the Temporal Score. The Base Score ``represents the intrinsic qualities of a vulnerability
that are constant over time and across user environments''~\mbox{\cite[p.~1]{cvssspecsv3.1}}.
The Environmental and Temporal Scores are optional and can be used to adapt the Base Score with respect to characteristics of a vulnerability that ``are unique to a user's environment''  or ``change over time''~\mbox{\cite[p. 1]{cvssspecsv3.1}}. 
The Base Score consists of eight metrics: Attack Vector (AV), Attack Complexity (AC), Privileges Required (PR), User Interaction (UI), Scope (S), Confidentiality (C), Integrity (I) and Availability (A). During evaluation, each metric is assigned a value (\Cref{tab:CVSSmetrics}) resulting in a vector string, e.g.: 
\begin{quote}
	CVSS:3.1/AV:N/AC:H/PR:L/UI:R/S:C/C:L/I:L/A:H
\end{quote}

Thus, the vector string consists of metric values assessed by an evaluator.
Each metric value has a corresponding numeric value. These numeric values are used in a formula to calculate the Base Score~\cite[p. 18]{cvssspecsv3.1}, which can range from 0.0 to 10.0, where 0.0 means no severity (None) and 10.0 means the highest severity (Critical) (see \Cref{tab:cvss-severity}).

FIRST published three documents for CVSSv3.1: Specification~\cite{cvssspecsv3.1} describes the scoring process, the metrics and the formulas; User Guide~\cite{cvssuserv3.1} provides scoring guidelines; Examples~\cite{cvssexamplev3.1} lists 29 vulnerabilities with explanation of their CVSS assessment.
According to FIRST, CVSS produces a numerical score that reflects severity of a vulnerability and can be used as input for vulnerability management. CVSS should not be used as the only input for ranking threats, as other factors such as monetary losses, threat of injury or public sentiment should also be considered. FIRST explicitly emphasizes  that CVSS measures severity and should not be used to measure risk.

The US-based National Vulnerability Database (NVD) is widely used within the CVSS community. For each vulnerability, it contains its CVE (Common Vulnerabilities and Exposures, a unique vulnerability ID) number, a short description, a CVSS evaluation, the CWE (Common Weakness Enumeration) identifier and advisory information. MITRE\footnote{\url{https://cve.mitre.org/}} provides the CVEs, whereas CVSS scores are contributed by NVD. CWE  reflects the type of vulnerability, e.g., Cross-site Scripting is CWE-79.

\begin{table}[h]
	\begin{center}
		\begin{tabular}{l l}
			\toprule
			Severity & Base Score \\
			\midrule
			None & 0.0 \\
			Low & 0.1 - 3.9 \\
			Medium & 4.0 - 6.9 \\
			High & 7.0 - 8.9 \\
			Critical & 9.0 - 10.0 \\
			\bottomrule
		\end{tabular}
		\vspace*{0.5em}
		\caption{Severity of the CVSS Base Scores~\cite[p.~21]{cvssspecsv3.1}.}
		\label{tab:cvss-severity}
	\end{center}
\end{table}

\subsection{Related Work}
\label{sec:related-work}

Different CVSS issues have been investigated and discussed within the security community (see \Cref{tab:cvss-relatedwork}).

\begin{table*}
	\begin{center}
		\begin{tabularx}{\linewidth}{l c c c c c c c c c}
			\toprule
			Paper & CVSS & Content & Year  & Partici- & Vulnera- &  Work & Working & Knowledge of & CVSS\\
			& Version & & & pants & bilities & Experience & Context & CVSS Documents & Attitude\\
			\midrule
			Scarfone, Mell~\cite{2009-scarfone} & 1, 2 & NVD Analysis & 2009 & - & 11,012 & $\medcircle$ & $\medcircle$ & $\medcircle$ & $\medcircle$\\
			Gallon~\cite{2011-gallon} & 2 & NVD Analysis & 2011 & - & 40,026 & $\medcircle$ & $\medcircle$ & $\medcircle$ & $\medcircle$\\
			Holm, Afridi~\cite{2015-holm} & 2 & Rater Accuracy & 2015 & 384 & 2131 & $\medblackcircle$ & $\medcircle$ & $\medcircle$ & $\medcircle$\\
			Klinedinst~\cite{2015-klinedinst} & 3 & Critique & 2015 & - & - & $\medcircle$ & $\medcircle$ & $\medcircle$ & $\medcircle$\\
			Spring et al.~\cite{2018-spring} & 3 & Critique & 2018 & - & - & $\medcircle$ & $\medcircle$ & $\medcircle$ & $\medcircle$\\
			Fall, Kadobayashi~\cite{2019-fall} & 2, 3 & NVD Analysis & 2019 & - & 40 & $\medcircle$ & $\medcircle$ & $\medcircle$ & $\medcircle$\\
			Allodi et al.~\cite{2020-allodi} & 3 & Rater Accuracy & 2020 & 73 & 30 & $\medblackcircle$ & $\medcircle$ & $\medcircle$ & $\medcircle$\\
			Spring et al.~\cite{2021-spring} & 3 & Critique & 2021 & - & - & $\medcircle$ & $\medcircle$ & $\medcircle$ & $\medcircle$\\
			\midrule
			This study & 3.1 & Scoring Consistency & 2023 & 196 & 10 & $\medblackcircle$ & $\medblackcircle$ & $\medblackcircle$ & $\medblackcircle$\\
			\bottomrule
		\end{tabularx}
		\vspace*{0.5em}
		\caption{Overview of related work that focused on CVSS. Related works that developed alternatives to CVSS or discussed CVSS in passing are not included in the table.}
		\label{tab:cvss-relatedwork}
	\end{center}
\end{table*}

\subsubsection*{Critique on CVSS}

Spring et al.~\cite{2018-spring} state that CVSS is often misused in companies, as the Base Score is used directly for risk rating, though it should be used for severity rating. Another problem is that the formula to compute the Base Score is not ``empirically or theoretically justified''~\cite[p. 2]{2018-spring}. The authors 
suggest conducting user studies to work out problematic aspects of CVSS in an evidence-based way. In a follow-up work, Spring et al.~\cite{2021-spring} criticize CVSS failure to account for context or consequences for human life, e.g., when assessing vulnerabilities of medical devices,
as this is not reflected by current CVSS metrics.
Klinedinst~\cite{2015-klinedinst} discusses the Scope metric as a useful addition in CVSSv3 and criticizes its lack of impact on the Base Score when choosing between S:U and S:C in some specific cases.

\subsubsection*{User Studies on CVSS}

Holm and Afridi~\cite{2015-holm} investigated the accuracy of CVSSv2 scores.
The participants of their survey assessed 3 fixed and 7 randomly selected vulnerabilities from NVD.
As a result, 38\% of evaluations differed in severity from NVD scores. The authors suggest revisions and additions to CVSS, e.g., environmental information and vulnerability scope. The latter was introduced in CVSSv3.0.

Allodi et al.~\cite{2020-allodi} measured   the accuracy of CVSSv3.0 assessments by evaluators with different IT security knowledge.
Security experts, information security students and computer science students evaluated 30 vulnerabilities randomly selected from a list  of 100 vulnerabilities used by FIRST to develop the CVSS formula. The authors compared the CVSS metric values assigned by the participants to the NVD values. However, they excluded the Scope metric ``[...] because there is a debate even inside the SIG expert group [...], due to the difficulty of correctly identifying its value even by CVSS’s own designers''~\cite[p. 6]{2020-allodi}. Computer science students were less accurate than the other two user groups, but the latter also produced a broad range of scores.

\subsubsection*{Investigation of NVD scores}
\label{sec:NVD}

NVD scores have been widely analyzed in the past years. 
Thus, Scarfone and Mell~\cite{2009-scarfone} analyzed the distribution of CVSSv1 and CVSSv2 scores within the NVD and showed that CVSSv2 scores provide a greater score diversity compared to CVSSv1.
Gallon~\cite{2011-gallon} investigated the distribution of CVSSv2 metrics  for vulnerabilities published by NVD between 1999 to 2009 and showed that most vulnerabilities are evaluated with medium Base Scores, which makes it hard to prioritize them.
Fall and Kadobayashi~\cite{2019-fall} analyzed the distribution of CVSSv3.0 assessments of the NVD concerning ``rock star'' vulnerabilities\footnote{Vulnerabilities that received a lot of media attention, e.g., Meltdown~\cite{2018-lipp-meltdown} or Spectre~\cite{2018-kocher-spectre}}. According to the authors, most of these vulnerabilities were rated with a lower severity than expected.
Several works uncovered poor information quality and accuracy of NVD entries and developed systems for improvement~\cite{2021-kuehn, 2020-anwar, 2019-dong}.

\subsubsection*{CVSS Alternatives}
Wang et al. \cite{2011-wang} defined new CVSS metrics that, e.g., reflect the OS type of the vulnerable system to developed an improved scoring system based on CVSSv2.
Spring et al. \cite{2020-spring} developed SSVC\footnote{Stackeholder-Specific Vulnerability Categorization}, which is a decision tree based scoring system built upon CVSSv3.0.
Zeng et al. \cite{2022-zeng} presented LICALITY, which uses neural network and program language processing techniques to assess risk of vulnerabilities combining criticality and likelihood of exploitation.
Ganin et al. \cite{2020-ganin} proposed a decision-analysis based framework for quantifying vulnerabilities, also considering consequences, risk assessment and management.
Chen et al. \cite{2019-chen} defined a framework for  exploitation prediction using Twitter resulting in classification and regression forecast models.
Alperin et al. \cite{2019-alperin} used natural language processing techniques to develop a vulnerability assessment system that also includes availability of exploits.

\subsubsection*{Research Gap}
The goal of our study is to evaluate the consistency of CVSSv3.1 scoring. In contrast to Allodi et al., we include the Scope metric.
Furthermore, we systematically select widespread vulnerabilities and problematic metrics to be investigated based on experience of longtime CVSS users, whereas other works selected vulnerabilities randomly. We also investigate for the first time users' attitudes to CVSS, and whether the daily routine of CVSS assessments and attitudes influence the scoring. We are also the first to examine whether evaluations of the same person change over time.
\section{Development of Research Questions}
\label{sec:development-research-questions}

To gain insight into problematic aspects of CVSS and develop research questions, we first conducted a discussion session with three CVSS experts. Thereafter, we validated discussion results and gathered more expert opinion in a preliminary survey with 18 experienced CVSS users.

\subsection{Discussion with CVSS Experts}
\label{sec:discussion-session}
In June 2020 we held a discussion session with three participants who conduct CVSS evaluations regularly since 2013-2014. These participants are referred to as \emph{discussion experts} in the following. The discussion experts were recruited through personal contacts. One of them conducts external CVSS evaluations (self-employed) and two conduct internal CVSS evaluations (one in a multinational automotive corporation, one in a large industrial manufacturing company). Two have a PhD and one has Master's degree in computer science. All are male, one is in 30-40 and two are in 40-50 age range. The experts were not compensated.
Three researchers made notes during the session, as recording was not allowed. Thereafter, we extracted, discussed and combined the main topics and insights.

The experts described their average working day and their experience with CVSS. Further, they were asked to discuss aspects of CVSS and vulnerability types where ambiguities and disagreements often occur. Following the discussion, we chose 10 initial vulnerabilities for further investigation in an expert survey (see \Cref{sec:preliminary-survey}). Due to space restrictions, these vulnerabilities are not presented in detail. We briefly outline results from the discussion and the survey in the end of \Cref{sec:preliminary-survey}. Subsequently, in \Cref{sec:implications} we elaborate on what we learned from the two preliminary studies and present the implications for the main study.

\subsection{Survey with CVSS Experts}
\label{sec:preliminary-survey}

We conducted an online survey to collect additional expert views on problematic CVSS aspects and to evaluate selected candidate vulnerabilities. Our recruiting was targeted at experienced CVSS users who regularly use CVSS in their daily work.
The participants were first asked to describe (via free text) possible issues with CVSS. Next, they assessed 10 candidate vulnerabilities using FIRST Online Calculator\footnote{\url{https://www.first.org/cvss/calculator/3.1}}. The order of the vulnerabilities was randomized. Additionally, the participants rated description quality, familiarity with the software and suitability for CVSS evaluation for each vulnerability on a 5-point Likert scale. They could also state their comments for each vulnerability as free text, which were later qualitatively coded by two researchers independently, and then discussed in a team meeting.

The preliminary study ran for three weeks in fall 2020. We used personal contacts and snowballing to recruit 18 participants. They have been assessing vulnerabilities using CVSS for 5 years on average, and most considered themselves advanced or expert CVSS user. Nearly all participants identified as male with an average age of 35 years. Their industry sectors were distributed over manufacturing, communication technology, healthcare, IT consulting and academic institutions, where most of them worked as information security managers, product security managers or members of incident response teams. The participants were not compensated, but expressed great interest in the topic and high motivation to help.

Participation in the study took 40 minutes on median. The free-text question about CVSS issues was asked before the vulnerability assessment to avoid priming.
The vulnerabilities were presented in randomized order to counteract possible fatigue. Additionally, at the end of the survey participants were asked to rate control statements, such as ``I have answered the questions carefully''. The result showed no abnormalities. Thus, although the survey took a long time, survey fatigue was likely avoided due to high commitment and professionalism of the participants.

Problems with CVSS stated in the free-text fields of the survey were quite similar to those from the discussion. Discussion experts and survey participants described CVSS evaluations  as  heavily subjective. Furthermore, they identified metrics that often lead to discussions. For example, discussion experts as well as 8 survey participants stated that the metric Scope is difficult to evaluate. 
Another problem, mentioned by discussion experts and 7 survey participants, is the difficulty to differentiate between AV:N and AV:A for the metric Attack Vector. 
The quantitative evaluation of the CVSS assessments showed that evaluations of many metrics were not consistent. Descriptions of vulnerabilities were mostly rated as sufficient and understandable, but free text comments showed that some details were missing. Therefore, we decided to replace some vulnerabilities and slightly rephrased some of the descriptions to make them more understandable for the main study.

\subsection{Implications for Study Design}
\label{sec:implications}

Preliminary expert studies revealed some aspects of CVSS that were not apparent to us from literature research and analysis of CVSS documentation. These aspects influenced our study design and are presented below.

\subsubsection*{(1) Number of evaluated vulnerabilities per participant should be low}
In the expert survey, all participants evaluated 10 vulnerabilities, taking on average 4 minutes per vulnerability. Several of them commented that they felt depleted afterwards, and suggested to reduce the number of vulnerabilities. Since we planned further questions in the main study besides the assessments, we decided to provide 4 vulnerabilities per person to avoid fatigue.

\subsubsection*{(2) Investigated CVSS metrics should be largely context-independent} 
The preliminary survey revealed that CVSS metrics Attack Complexity, Privileges Required and Impact Metrics (Confidentiality, Integrity and Availability) strongly depend on the context of the vulnerability, such as typical installation of the system or software knowledge. For example, one participant stated: \emph{``I was not sure about the level of the required privileges [$\dots$] as I am not familiar with this application framework''}. Another participant commented, referring to missing system knowledge to rate the Impact Metrics: \emph{``[$\dots$] unclear which exact data gets endangered''}.

In contrast, metrics Attack Vector, User Interaction  and Scope received a low number of context-dependent comments. For example, although many participants found Scope difficult to rate, most of these difficulties referred to their low conceptual understanding of this metric. In another example, it is largely independent of the software whether a user has to be involved into a successful compromise (User Interaction). We note that in the main survey with 196 participants, metrics AV, UI and S received comments on being unclear to rate due to the lack of context 8, 3 and 9 times, respectively. At the same time, AC, PR and Impact Metrics received such comments 18, 16 and 67 times, respectively.

We decided to investigate the less context-dependent metrics first, as fitting all metrics with their context into descriptions of the vulnerabilities would make the descriptions substantially longer. This would make the time spent per vulnerability longer as well, and the number of rated vulnerabilities per user even lower than outlined in point~(1). 

\subsubsection*{(3) We should aim at problematic metrics for widespread vulnerability types}
Previous user studies~\cite{2020-allodi,2015-holm} selected vulnerabilities randomly. However, discussion experts indicated vulnerabilities that they evaluate on a daily basis, and where some \emph{particular} metrics often lead to discussions with colleagues. We call such vulnerabilities ``widespread'', and such metrics ``problematic''.
To determine which of these vulnerabilities may be especially important, we analyzed the Examples document, where CVSS assessments of specific vulnerabilities are presented~\cite{cvssexamplev3.1}. Furthermore, we also consulted the 2022 CWE Top 25 list, where the currently most common and impactful vulnerabilities are listed. To create this ranking, MITRE analyzed public vulnerability data from NVD and applied a scoring formula to create an order that incorporated both frequency and severity of vulnerabilities \cite{cwe-top25}.
Later, to test whether the chosen vulnerabilities were widespread in our sample, we asked participants in the main study how often they rated vulnerabilities of this type in their daily work (see \Cref{tab:matrix-questions}). Finally, we included the 2022 CWE Top 3 vulnerabilities, as well as some other vulnerabilities that were considered important in the expert studies.
\section{Research Questions}
\label{sec:research-questions}

We present research questions that originated from the expert discussion and were validated and adjusted using the results of the preliminary expert survey.

\subsection{Consistency of Evaluations of Problematic Metrics for Widespread Vulnerability Types (RQ1)}
\label{sec:metrics-vulns}

As described in \Cref{sec:implications}, we selected the metrics Attack Vector, User Interaction and Scope for further investigation. These metrics were described as problematic in expert studies for some common and impactful vulnerabilities. This led to the following research question:

\begin{quote}
\textbf{RQ1:} Are metrics Attack Vector, User Interaction and Scope inconsistently evaluated for some widespread vulnerability types? 
\end{quote}

In the following, we present metrics and vulnerability types that we investigate. \Cref{sec:vulnerabilities-groups} presents concrete vulnerabiltites (\Cref{tab:vulnerabilities}) and explains how they were selected and assigned to participants.

\subsubsection*{Attack Vector and Drive-by Download} 
When evaluating the metric Attack Vector it is often unclear what value, Network or Local, should be selected if the vulnerability concerns Drive-by Download attacks\footnote{Drive-by download vulnerabilities let a victim unknowingly download malicious code and execute it.}. When downloaded, the malicious code is usually locally saved on the PC, therefore one can argue for AV:L. However, the code needs to be accessed from the network, so one can argue for AV:N. This issue is discussed in depth in User Guide~\cite[p. 10]{cvssuserv3.1} and in Examples~\cite[p. 37]{cvssexamplev3.1}, which shows that this case is important and difficult. The expert studies also showed that rating AV:N or AV:L in this case is perceived as ambiguous. These vulnerabilities are currently also most common and impactful, as the most dangerous of them lead the 2022 CWE 25 list: CWE-787	Out-of-bounds Write.

\subsubsection*{Attack Vector and MITM} 
Choosing between AV:N and AV:A when rating MITM (man-in-the-middle) vulnerabilities is difficult according to discussion experts, which was confirmed by the expert survey. 
A reason for AV:A could be  the assumption that MITM attacks can only be carried out within the local broadcast domain, close to the target system. CVSS documentation instructs to score AV:N, as the attacker needs network level to access the communication channel~\cite[p. 33]{cvssexamplev3.1}.

\subsubsection*{User Interaction and XSS} 
In Reflected XSS (cross-site scripting) vulnerabilities, the attacker needs to convince the victim to click on a specific malicious link, which leads to a website where malicious code is executed. In Stored XSS vulnerabilities, the malicious code within a website is executed every time a user visits this site. In this vulnerability type, the attacker does not need to convince a victim directly, as the malicious code is executed for every user who visits the website. Therefore, one could argue to score UI:N. However, Examples instruct to score UI:R, as ``the victim needs to navigate to a web page on the vulnerable server [...].''~\cite[p. 25]{cvssexamplev3.1}. XSS vulnerabilities are Top 2 in the 2022 CWE Top 25 list.
Thus, we investigate if the evaluation of User Interaction is more consistent for Reflected XSS than for Stored XSS. 

\subsubsection*{Scope} 
In the expert survey the evaluations of Scope differed for most vulnerabilities, and 8 out of 18 participants explicitly called the Scope metric difficult to evaluate. 
The expert survey showed that Scope leads to inconsistencies especially for XSS, SQL Injection  and Drive-by Download, which are Top 1-3 in the 2022 CWE Top 25 list. Therefore, we focus on these vulnerabilities for Scope.

\subsection{Suitability of Security Deficiencies (RQ2)}
\label{sec:security-deficiencies}
Another possible problem identified by the expert studies concerns security deficiencies. These can facilitate an attack in combination with a vulnerability, but do not lead to a successful attack per se, whereas a vulnerability ``could be accidentally or intentionally exploited to damage assets.'' \cite[p.~24]{2011-gollmann}. If no damage occurs, CVSS severity should be rated as None~\cite[p.~9]{cvssspecsv3.1}. Yet, the discussion experts stated that in practice, security deficiencies are usually assigned CVSS scores of a very broad severity range.
This led to the following research question:

\begin{quote}
\textbf{RQ2:} Are security deficiencies considered suitable for CVSSv3.1 assessment by CVSS users?
\end{quote}
Following the discussion with experts, we analyze two security deficiencies:
 
\emph{(1) Missing HTTPOnly Flag} is a missing protection mechanism intended to make exploitation of XSS vulnerabilities more difficult. While enabling this flag can be beneficial, its absence does not mean that an application is vulnerable to XSS attacks. Yet, NVD contains several entries of this type~\cite{cve-2020-27658, cve-2014-9635}, and discussion experts stated that they often rate them.

\emph{(2) Banner Disclosure} means disclosing information on use of a particular web server and its version in HTTP responses~\cite{cve-2017-4013,cve-2020-3193}. This disclosure makes it easier for an attacker to search for known vulnerabilities in the product, although the disclosure alone will not cause any damage.

Our goal is to investigate whether these vulnerability types are more likely to be rated with severity None than other vulnerabilities, and to gather opinion of CVSS users on these issues. Because of this, we do not focus on specific metrics here, but on the severity of scoring. 

\subsection{Relation to Personal Factors (RQ3)}
\label{sec:personal-factors}
Allodi et al.~\cite{2020-allodi} found that with increasing years of work experience, accuracy of evaluations also increases. Thus, personal factors such as work experience might affect the consistency of ratings, which led us to the following research question.
\begin{quote}
\textbf{RQ3:} Are personal factors associated with the individual differences in CVSSv3.1 evaluations?
\end{quote}

We determined possible factors from related work, CVSS documentation and the expert discussion. 
In line with Allodi et al. we analyzed whether work experience and experience with CVSS has an influence on scoring. As the CVSS documentation
describes the correct usage of CVSS, we assume that knowledge and frequency of consulting these documents is related to consistency of evaluations. The discussion experts reported different working routines, e.g., which tools are used 
or how many evaluators participate in one CVSS assessment. Therefore, we tested whether differing procedures of CVSS evaluations may influence consistency of evaluations. The discussion experts also reasoned that the evaluators' attitude to CVSS (e.g., is it useful and easy to use?) may influence their evaluations.

\subsection{Attitudes to CVSS (RQ4)}
Attitudes to CVSS have not been investigated so far. As the system received a lot of critique over the years (\Cref{sec:related-work}), it seems to be important to know how it is perceived by its users. For example, whether it is seen more as a useful tool or as a burden and cause for discussions when evaluating vulnerabilities. This led to the following research question:
\begin{quote}
	\textbf{RQ4}: What is the attitude of evaluators towards CVSSv3.1?
\end{quote}
In the survey, we collected participants' ratings of specific statements, such as whether CVSS is a useful or whether scores by different users often differ. Participants were also asked to share their thoughts about CVSS as free text.

\subsection{Consistency of Evaluations Over Time (RQ5)}
During the expert discussion, one participant mentioned a case where they re-evaluated the same vulnerability after some time, without any additional information added between the two evaluations. The resulting scores differed, which led to a discussion of consistency of participants' own evaluations over time. Generally, they assumed that they would not evaluate consistently, which led us to the following research question. 

\begin{quote}
\textbf{RQ5:} Are CVSSv3.1 evaluations of the same evaluator consistent over time?
\end{quote}
To answer RQ5, we conducted a follow-up study nine months after the main study in which participants reassessed vulnerabilities from the main study. 
\section{Study Design and Data Analysis}
\label{sec:methods}

\subsubsection*{Selection of Vulnerabilities and Group Assignment}
\label{sec:vulnerabilities-groups}

According to RQ1 (\Cref{sec:metrics-vulns}) and RQ2 (\Cref{sec:security-deficiencies}), and to keep the number of rated vulnerabilities low, we needed to select vulnerabilities of the following types: 

\begin{itemize}
	\item Two Drive-by Downloads of the CWE-787 (Out-of-bounds Write) type, one with AV:N, and one with AV:L
	\item One Reflected XSS and one Stored XSS vulnerability concerning the same software (to be comparable)
	\item One MITM vulnerability
	\item One SQL Injection vulnerability
	\item One Banner Disclosure and one Missing HTTPOnly Flag security deficiency
\end{itemize}

The choice was first made by three researchers independently, then the resulting vulnerabilities were discussed and selected. CVSS 3.1 Examples document contains two well described Drive-by-Download vulnerabilities~\cite[p. 26 and p. 36]{cvssexamplev3.1}, and other vulnerabilities were selected directly from NVD. To keep the number of CVSS assessments per participant feasible, the eight selected vulnerabilities were divided into two groups (\Cref{tab:vulnerabilities}). We assigned vulnerabilities of the same type to the same group, which results in both XSS vulnerabilities being in Group 1 and both Drive-by Download vulnerabilities in Group 2. Each participant also evaluated one security deficiency. Full descriptions of the vulnerabilities are stated in Appendix~\ref{sec:appendix-questionnaire-vuln-descriptions}.

\subsubsection*{Questionnaire Setup} The structure of the questionnaire is depicted in \Cref{fig:mainstudy-questionnaire}. The full questionnaire is presented in Appendix \ref{sec:appendix-questionnaire-mainstudy}. Participants first provided informed consent and were asked whether they are currently assessing vulnerabilities using CVSS, as our study targeted only those participants. 
Eligible participants were first asked about their work experience: their position, years in their current job, years of working with CVSS and areas of expertise, and then self-assessed their CVSS expertise. 
Next questions were concerned with daily work using CVSS, such as: which tools and documents are used, how CVSS is used for vulnerability management, how many persons are usually involved in a CVSS assessment, and how long one evaluation takes on average. Next, participants were asked to assess their knowledge of Specification~\cite{cvssspecsv3.1}, User Guide~\cite{cvssuserv3.1} and Examples~\cite{cvssexamplev3.1}, and when they have last consulted these.

\begin{figure*}
	\begin{center}
		\includegraphics[width=0.8\textwidth]{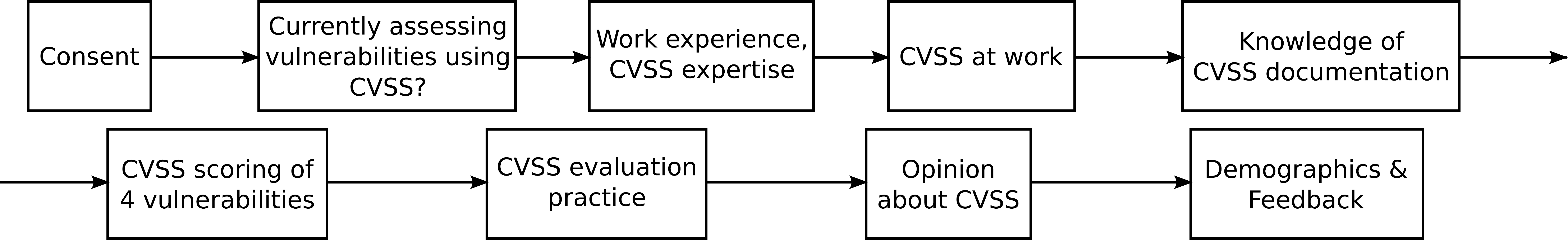}
	\end{center}
	\caption{Structure of the main study.}
	\label{fig:mainstudy-questionnaire}
\end{figure*}

Then each participant was randomly assigned to one of the two groups in \Cref{tab:vulnerabilities} and received the corresponding four vulnerabilities in randomized order. After each assessment, the participants indicated on a 5-point Likert scale their scoring confidence, knowledge of the software, clarity of the description, suitability of the vulnerability for CVSS assessment and whether they often rate similar vulnerabilities in their daily work.
The participants were also asked to state their assumptions for the assessment and any additional comments as free text.

After the CVSS assessment, we asked the participants how they would evaluate some specific cases from CVSS documentation (Appendix~\ref{sec:unklarheitenFragen}). In this way, we measured how closely they adhere to the documentation. Next, we measured participants' attitudes to CVSS, such as whether CVSS is considered useful and low-effort. The survey concluded with demographic questions.
The questionnaire contained three attention checks. Participants who failed two or more of them were excluded from the analysis.

The survey was tested by 10 CVSS evaluators. This did not result in any questions or changes to the content.

\subsubsection*{Ethics and Recruitment}
The study was approved by the data protection office of the Friedrich-Alexander-Universität Erlangen-Nürnberg (FAU).
The participation was voluntary and could be aborted anytime without consequences. The participants were not compensated. They provided informed consent, and for demographics a ``Prefer not to say'' option was pre-selected for each question.
 For recruitment, we advertised the study on the web page of the IT Security Infrastructures Lab of the FAU and through private contacts.
We also used mailing lists SecurityMetrics and FullDisclosure and posted invitations in various Reddit areas connected to vulnerability assessment. Two articles about CVSS on Heise.de~\cite{heiseCVSS} and The Register~\cite{theregister} also advertised the survey.
Additionally, we asked every participant to forward the invitation link to other suitable people. The survey was conducted in winter 2020/21. Participation in the study took an average of 30 minutes.

\subsubsection*{Quantitative Data Analysis}
\label{sec:quant-analysis}
Analyses were performed using Stata SE 14.2 assuming a significance level of \mbox{$p<0.05$}. 
When presenting regressions, $t$-statistics for each effect are displayed in parentheses; significance levels are showed by superscripts: $^{*} p < 0.05, ^{**} p < 0.01,$ $^{***} p < 0.001$.

Ground truth for evaluations was created by considering similar vulnerabilities in CVSS Examples \cite{cvssexamplev3.1}. Evaluations of the security deficiencies were not considered, since we only consider whether they are more frequently rated as None severity than other vulnerabilities (\Cref{sec:security-deficiencies}). A post-hoc comparison of the average number of correct responses per group showed that they differ  slightly (Group 1: $\sigma = 2.7, \mu = 1.3$; Group 2: $\sigma = 3.2, \mu = 1.1$)\footnote{$\sigma = $ mean, $\mu = $ standard deviation} but significantly (ANOVA: $F(1,194) = 6.09, p < 0.05, R^2 = 0.03$), meaning that tasks in Group 1 were more difficult than tasks in Group 2. We decided to treat this difference as a measurement error, as our research questions do not consider task difficulty as factor. To control for this difference, a standardization by group mean was performed~\cite[p.~101]{2018-spiegel}: each measurement ($x_i$) was compared with the group mean ($x$) and divided by the standard deviation of the variable ($s$): $z_i=\frac{(x_i-x)}{s}$. This transformation resulted in a continuous dependent variable with a minimum of -2.1 and a maximum of 1.7, such that linear regression was applied. A comparison after standardization showed, as expected, no statistically significant difference between the groups. 
 
We used an exploratory approach to data analysis~\cite{1977-tukey}. The first step was to check for each variable of interest whether there was a relationship with the dependent variable using stepwise regression models~\cite{2002-thayer}. Then, all variables for which there was a significant correlation were selected for the final regression analysis, and supplemented by other variables that the researchers considered to be of particular interest, even if no significant correlation was found for them. The advantage of using multiple regression is that the influence of one independent variable can be shown while controlling for each of the other variables. 

\begin{table}
	\begin{center}
		\begin{tabular}{l c c}
			\toprule
			Vulnerability & CVE & Metric:GT\\
			\midrule
			Group 1: $n=97$ & &\\
			Reflected XSS & CVE-2019-20512~\cite{cve-2019-20512} & UI:R, S:C\\
			Stored XSS & CVE-2020-13145~\cite{cve-2020-13145} & UI:R, S:C\\
			SQL Injection & CVE-2020-3184~\cite{cve-2020-3184} & S:U\\
			Banner Disclosure & CVE-2020-3193~\cite{cve-2020-3193} & --\\
			\midrule
			Group 2: $n=99$ & &\\
			Adobe Acrobat & CVE-2009-0658~\cite[p. 26]{cvssexamplev3.1} & AV:L, S:U\\
			Google Chrome & CVE-2016-1645~\cite[p. 36]{cvssexamplev3.1} & AV:N, S:U\\
			MITM MyPalette & CVE-2020-5523~\cite{cve-2020-5523} & AV:N\\
			HTTPOnly & CVE-2020-27658~\cite{cve-2020-27658} & --\\
			\bottomrule
		\end{tabular}
		\caption{Groups of vulnerabilities for the main study. Each participant was randomly assigned to evaluate vulnerabilities from one of the groups, presented in randomized order. GT = Ground Truth (correct value of the metric). For security deficiencies (Banner Disclosure and HTTPOnly), we did not consider ``correctness'' of ratings (see \Cref{sec:security-deficiencies}).}
		\label{tab:vulnerabilities}
	\end{center}
\end{table}
To measure the level of agreement between participants, we use Finn's coefficient~\cite{1970-finn, 2021-gwet}. We used the following classification levels: very poor, for values between 0.00 and 0.20; poor
(between 0.21 and 0.40); fair (between 0.41 and 0.60); moderate (between 0.61 and 0.80); and substantial (between 0.81 and 1.00) agreement.

\subsubsection*{Qualitative Data Analysis}
There were two kinds of free-text answers: First, the participants could annotate difficulties or assumptions after each assessed vulnerability, and second, at the end of the survey they could share additional thoughts on CVSS.  
For the 311 answers concerning vulnerabilities, we performed intercoder agreement using Cohen's Kappa $\kappa$~\cite{1960-cohen}, as these comments had the potential for ambiguous interpretation~\cite{mcdonald2019reliability}. One researcher first read all comments and identified the first set of codes. These codes were discussed with a second resarcher and converted into a codebook that was used by both researchers for the initial coding of 40 answers (5 for each of the eight vulnerabilities). They then discussed the disagreements, adjusted the codebook and subsequently coded all  answers. For 11 out of 21 codes the agreement was substantial ($\kappa > 0.75$), and for 7 it was moderate ($\kappa > 0.4$). The remaining 3 codes were used as a catch-all category for unclear statements and contained 13 entries. Finally, all disagreements were discussed, and full agreement could be reached.

The 55 additional comments on CVSS were analyzed without an intercoder agreement.  Here we leaned on McDonald et al’s exclusion criterion, ``when coding requires little interpretation'' ~\cite[p. 72:3]{mcdonald2019reliability}.
Two researchers first read all comments and identified codes that were discussed to create a codebook. Then this codebook was used by one researcher to code all comments, which was in turn subsequently reviewed by another researcher. As no ambiguities or disagreements emerged when the two researchers discussed selected codes, we deem McDonald et al’s criterion fulfilled. After the coding, both researchers summarized the free-text answers for each code independently. These summaries were discussed and converted into a single summary.

\subsubsection*{Participants}
\label{sec:mainstudy-participants}
In total 207 people completed the main survey. Ten were excluded because they did not pass the attention tests, and one was under 18, which leaves 196 valid answers.
\Cref{tab:demographics} presents participants' demographics. Nearly all identified as male, 7 as  female and one as diverse.
On average, the participants were 38 years old, the youngest was 19, and the eldest 63. The majority were from Europe (mostly Germany and the UK) and the USA. Most had a Masters's or a Bachelor's degree and were working as an employee or civil servant. A third of the participants were employed at companies with 10,000 or more employees, and a third were working at companies with less than 500 employees. As the economic sector of their organization, 45\% 
indicated information and communication, around 10\% provided the sector for professional, scientific and technical activities, and manufacturing was also indicated by 10\%.
Most participants self-assess their expertise in CVSSv3.1 at intermediate or advanced level. On average, they have been assessing vulnerabilities using CVSS for 6.5 years.

\begin{table}
	\begin{center}
		\begin{tabularx}{0.8\linewidth}{l|X X}
			\toprule
			& $N$ & $\%$ \\
			Total & 196 & 100,0 \\\hline
			Male & 170 & 86.7\\
			Female & 7 & 3.6\\
			Diverse & 1 & 0.5\\
			N/A & 18 & 9.2\\\hline
			18-29 years & 30 & 15.3\\
			30-49 years & 110 & 56.1\\
			50-60 years  & 19 & 9.7\\
			Above 60 years & 4 & 2.0\\
			N/A & 33 & 16.8\\\hline
			Germany & 48 & 24.5\\
			United Kingdom & 25 & 12.8\\
			Other European country & 46 & 23.5\\
			USA & 38 & 19.4\\
			Asia & 6 & 3.1\\
			Other country & 6 & 3.1\\
			N/A & 27 & 13.8\\\hline
			No academic education & 17 & 8.7\\
			Bachelor's degree & 73 & 37.2\\
			Master's degree & 85 & 43.4\\
			Ph.D. & 6 & 3.1\\
			Other & 7 & 3.6\\
			N/A & 8 & 4.1\\\hline
			Employee, civil servant & 165 & 64.2\\
			Self-employed, freelancer & 12 & 6.1\\
			Student & 1 & 0.5\\
			Other & 9 & 4.6\\
			N/A & 9 & 4.6\\\hline
			1-3 years using CVSS & 56 & 28.6\\
			4-6 years using CVSS & 65 & 33.2\\
			7-10 years using CVSS & 50 & 25.5\\
			More than 10 years using CVSS & 25 & 12.8\\\hline
			None or basic CVSSv3.1 expertise & 39 & 19.9\\
			Intermediate CVSSv3.1 expertise & 60 & 30.6\\
			Advanced or expert CVSSv3.1 expertise & 97 & 49.5\\
			\bottomrule
		\end{tabularx}
		\vspace*{0.5em}
		\caption{Overview of the participants' demographics, CVSS experience and self-assessed CVSS expertise.}
		\label{tab:demographics}
	\end{center}
\end{table}
\section{Results}
\label{sec:results}
Our approach was to investigate the consistency of evaluations of problematic metrics for widespread  vulnerabilities. Most participants agreed that the descriptions of the vulnerabilities were clear and understandable, and that they often rate similar vulnerabilities in their daily work~(\Cref{tab:matrix-questions} in Appendix \ref{sec:appendix-table-matrix-questions}).

\subsection{Evaluation Consistency (RQ1)}
\label{sec:results-reliability-metrics}

\begin{figure}
	\begin{center}
		\includegraphics[width=0.45\textwidth]{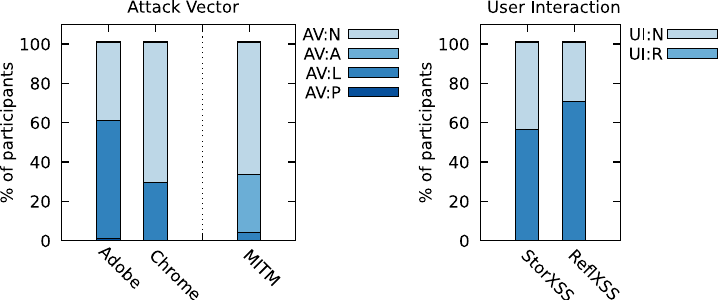}
	\end{center}
	\caption{Evaluations of the Attack Vector and User Interaction metric.}
	\label{fig:results-av-ui}
\end{figure}

\begin{figure}
	\begin{center}
		\includegraphics[width=0.25\textwidth]{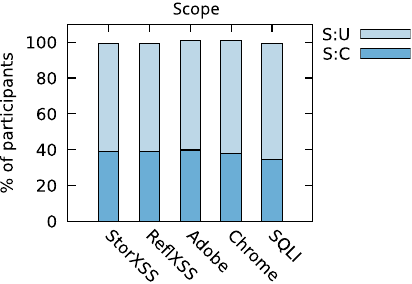}
	\end{center}
	\caption{Evaluations of the Scope metric.}
	\label{fig:results-s}
\end{figure}

\subsubsection*{AV \& Drive-by Download}
For Google Chrome, around 70\% of the participants chose AV:N, whereas 30\% chose AV:L (\Cref{fig:results-av-ui}). Finn's coefficient is $0.33$ $(F(Inf,98)=1.49,~p=0.0049)$, indicating poor agreement. P101 summarized their considerations: \emph{``I think the attack vector in CVSS is unsuitable for this kind of vulnerability. Is this network since it is delivered through the network [...]? Or local since Chrome obviously doesn't open a network port that is attacked here''.} Choosing AV:N matches the official guidelines for CVSS~\cite[p. 37]{cvssexamplev3.1}.

For Adobe Acrobat, around 60\% of the participants chose AV:L, whereas 40\% chose AV:N. Finn's coefficient indicates poor agreement: $0.21$ $(F(Inf,98)=1.26,~p=0.0647)$. Free-text comments provide insight into this difference: \emph{``[The PDF] comes from the network,''} P112 stated, whereas P182 justified: \emph{``[...] the user gets the document via e.g. email [...], but not counting this as an AV:N, as this is not tied to the network stack directly''.} According to Examples, the vulnerability should be scored with AV:L, as this vulnerability is ``a flaw in the local document software that is triggered by opening a malformed document,'' \cite[p. 27]{cvssexamplev3.1}. However, Examples also recommends to chose AV:N if the victim could load the file either via network or from local media. Additionally, User Guide instructs to ``rate vulnerabilities where the malicious data is received over a network by one component, then passed to a separate component with a vulnerability''~\cite[p. 16]{cvssuserv3.1} as AV:L. Thus, the guidance is spread across different documents and might be unclear.

\subsubsection*{AV \& MITM}
Approximately 60\% of the participants chose AV:N, whereas nearly 30\% chose AV:A (\Cref{fig:results-av-ui}). For example, P152 explained that a MITM attack is \emph{``[...] most likely to occur on an adjacent network [...] rather than across the internet''}. However, the MITM vulnerability could also be exploited across the network, which makes scoring AV:N also appropriate. Finn's coefficient is $0.746$ $(F(Inf,98)=3.93,~p=4.17^{e-15})$. Although the distribution of the ratings here is similar to the Drive-by Download vulnerabilities, Finn's coefficient indicates moderate agreement, whereas it indicated poor agreement for Drive-by Downloads. This difference is due to the fact that the Finn's coefficient should be calculated on an ``equal appearing interval scale''~\cite[p. 76]{1970-finn}. Since AV:N and AV:A follow each other and AV:N and AV:L have a rating ``in between'' (AV:A), the Finn's coefficient calculates a higher agreement value.
Thus, there are noticeable inconsistencies in evaluating AV for MITM vulnerability types, and it seems that scoring of AV:A versus AV:N needs further clarification.

\subsubsection*{UI \& XSS}

For Reflected XSS, 76\% of the participants chose UI:R. Finn's coefficient is $0.247$ $(F(Inf,96) = 1.33,~p = 0.0335)$, which indicates a poor agreement. However, for Stored XSS, only 58\% of the participants chose UI:R (see \Cref{fig:results-av-ui}). Finn's agreement coefficient is $0.0206$ $(F(Inf,96)=1.02,~p=0.462)$, which indicates that there is almost no agreement, and the ratings are largely inconsistent. 
In the free-text answers, the participants argued for both cases, UI:R and UI:N for Stored XSS. 
P76 summarized: \emph{``UI:R vs. UI:N is often debated for Stored XSS. At the moment I'm in the UI:R camp because I consider the victim's visit the affected page as already a User Interaction''}.
Compared to Stored XSS, evaluating Reflected XSS seems to be much clearer. The reason behind this might be that for Reflected XSS the common knowledge is that the victim needs to click on a link to trigger the malicious code. However, for Stored XSS the victim also ``needs to navigate to a web page on the vulnerable server [...],'' \cite[p. 25]{cvssexamplev3.1}, which indicates UI:R. Thus, a guideline exists, but seems to be unknown.
We used a proportion test to analyze whether User Interaction is more consistent for Reflected XSS than for Stored XSS. The difference is significant ($z = 19.850; p < 0.05$) with Cohen's $h=0.35$ (medium effect size)~\cite{1977-cohen}. 

\subsubsection*{Scope}

Each of the vulnerabilities for Scope was evaluated nearly with the same relation: around 60\% of the participants chose S:U, whereas 40\% chose S:C (\Cref{fig:results-s}). All Finn's coefficients were far below $0.2$, which indicates a very poor agreement: $0.0369$ $(F(Inf,96)=1.04,~p=0.416)$ for Reflected XSS, $0.0369$ $(F(Inf,96)=1.04,~p=0.416)$ for Stored XSS, $0.0799$ $(F(Inf,96)= 1.09,~p=0.301)$ for SQL Injection, $0.0353$ $(F(Inf,98)=1.04,~p=0.42)$ for Adobe Acrobat and $0.0542$ $(F(Inf,98)=1.06,~p= 0.367)$ for Google Chrome. The free-text answers clarify this phenomenon: \emph{``I have never evaluated something with a scope change since S:C seems almost unachievable,''} P135 stated. The participants described Scope as unclear, confusing and hard to understand. P189 summarized: \emph{``Scope: If you ask 10 people for their opinion you get 10 coin tosses. [...] I spent some time trying to understand it before CVSS3.1 came out from docs, examples, and conversations with others but never understood it fully''}.
The  documentation provides guidance and examples for the Scope metric~\cite[p. 11]{cvssuserv3.1}, but they seem to be incomprehensible. 

\subsection{Security Deficiencies (RQ2)}

\begin{figure}
	\begin{center}
		\includegraphics[width=0.5\textwidth]{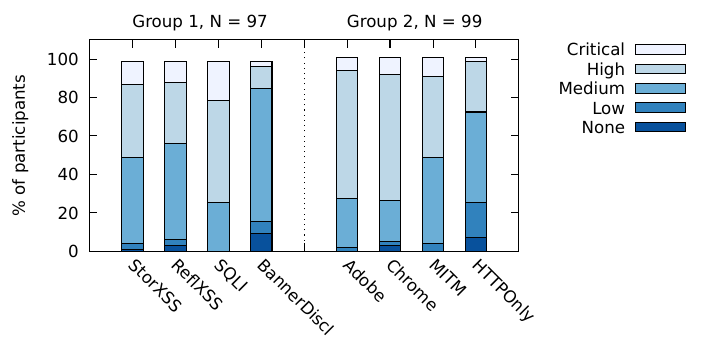}
	\end{center}
	\caption{RQ2: Severity distributions of the vulnerabilities. Security deficiencies Banner Disclosure and HTTPOnly are more frequently rated with None than other vulnerabilities.}
	\label{fig:overview-severity}
\end{figure}

\Cref{fig:overview-severity} provide an overview of the severity distributions for rated vulnerabilities. Banner Disclosure was mostly evaluated with Medium severity, but has the highest percentage of None ratings (9.2\% evaluations). Severity of \mbox{HTTPOnly} varied the most compared to the other vulnerabilities, where 7.1\% of participants assigned None severity. Considering that of the remaining vulnerabilities, three were not assigned None severity at all, and the percentage of None scores for the others is between 1\% and 3\%, Banner Disclosure and HTTPOnly stand out.
Additionally, several participants commented that they do not consider them as vulnerabilities, e.g.: \emph{``It's not a vulnerability [...], rather a missing mitigation and ignores best practices. CVSS 0 is appropriate''} (P100).
Other participants found the resulting scores too high: \emph{``I think it’s difficult to evaluate because server informations aren’t pretty sensitive informations [...] In my oppionin [sic] the only vulnerably point is confidentiality. To mark confidentiality with none would be result in a score of 0. This wouldn’t be representing the severity of this vulnerabilty [sic],''} (P91). P73 adjusted metric values until the resulting score seemed intuitively appropriate: \emph{``This calculation is forced to have Low score by arbitrarily selecting Physical attack vector [...]''}.
This shows that there is a need for the CVSS documentation to specify how security deficiencies should be handled.

\subsection{Relation to Personal Factors (RQ3)}

We first describe experience and working context of our sample, for the first time shedding light on everyday CVSS usage. Further we present a regression analysis of how these factors are associated with the correctness of ratings.

\subsubsection*{Experience and Documentation Knowledge}
\begin{figure}
	\begin{center}
		\includegraphics[width=0.45\textwidth]{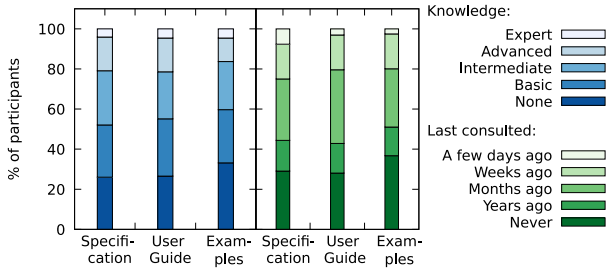}
	\end{center}
	\caption{Knowledge of CVSSv3.1 documentation and how recently the documents were consulted (N = 196).}
	\label{fig:results-documentation}
\end{figure}

On average, the participants have worked in IT security for 12 years and in their current position for 6 years. They have assessed vulnerabilities using CVSS for 6.5 years on average.
Approximately 60\% of participants see themselves on advanced or expert level in at least one of the following areas: network, web application or system software security.
Most participants self-assess their expertise in CVSSv3.1 and v3.0 at intermediate or advanced level, whereas the expertise in CVSSv2 is mostly on basic to intermediate level. FIRST offers online courses to get familiar with CVSS. Out of 196 participants, 11 attended the CVSSv3.1 and 13 the CVSSv3.0 FIRST course.
Over half of the participants declared their knowledge of the CVSSv3.1 documentation at none or basic level, whereas only 20\% indicated advanced or expert level (\Cref{fig:results-documentation}). Correspondingly, over 30\% never consulted the documentation, and around 50\% last consulted it years or months ago.

\subsubsection*{Working Context}
\label{sec:working-context}
\begin{table}[h]
	\begin{center}
		\begin{tabularx}{\linewidth}{X|l}
			\toprule
			\multicolumn{2}{c}{Question: The metric Availability reflects the availability of ...}\\
			\midrule
			Answers & $n$\\
			\midrule
			... performance and operation of the impacted component. & 74\\
			... data used by the impacted component. & 0\\
			... data, performance and operation of the impacted component. & 112\\
			... Other, please indicate: (text box) & 6\\
			... I don’t understand this question. & 4\\
			\bottomrule
		\end{tabularx}
		\caption{Example of a question concerning a specific case in CVSS documentation (N = 196). The first answer is correct according to documentation.}
		\label{tab:results-ambig-availability}
	\end{center}
\end{table}

Often there is no single default version for CVSS at work, but rather two or more, depending on the request of the customer. Most participants (60\%) use CVSSv3.1 for daily assessments, CVSSv3.0 is used by 38\% and CVSSv2.0 by 13\%. Furthermore, 73\% assess vulnerabilities using CVSS for their own organization, and 60\% evaluate scores for customers. The median time for a CVSS assessment is 5 minutes and the median number of CVSS assessments per week is 10. Base Scores of 55\% of the participants are always or often reviewed by coworkers. Only 6\% indicated that their Base Scores are never reviewed. On average, 10 evaluators are usually involved in a CVSS assessment.
During the evaluation process, 77\% of the participants use an online calculator, and 44\% use internal company-specific documents and software. Official CVSS documentation is used by approximately 25\%.
If an uncertainty occurs while evaluating, most of the participants usually consult a coworker; half of the participants do internet research or look for ratings of similar issues. 

A substantial 83\% of participants use CVSS for determining vulnerability severity, while 56\% utilize CVSS within their risk assessment procedures. 59\% declare to use CVSS for vulnerability prioritisation. Although our survey design doesn't allow us to determine whether CVSS is the primary factor in prioritisation decisions, the high prevalence of CVSS use across these facets underscores its fundamental role in vulnerability management strategies.

Besides the Base Score, there are also two optional scores: the Environmental and the Temporal Scores, which can be used to consider additional factors, e.g., if patches are available. The optional scores are only used by 55\% of the participants at work, whereas 70\% 
would like to adapt the Base Scores with at least one optional score. Using only the Base Score for vulnerability prioritization could lead to misallocated resources to mitigate supposedly high-severity vulnerabilities. Most participants are aware of this problem, but the usage of the optional scores is determined by work.

While analyzing CVSS documentation, we noticed that evaluators who are not familiar with these documents might choose different metric values than evaluators with detailed knowledge. We transferred seven of these cases into single-choice questions (\Cref{sec:unklarheitenFragen}) with one ``correct'' answer according to CVSS documentation. In this way we could investigate whether evaluators who provide ``correct'' answers have better performance.
For example, {``[...] the Availability metric speaks to the performance and operation of the service itself – not the availability of the data''~\cite[p. 9]{cvssuserv3.1}. This case leads to the question presented in \Cref{tab:results-ambig-availability}. Most participants (112 out of 196, 57\%) take data into account when evaluating the Availability metric, whereas 74 participants (38\%) only consider performance and operation of the service, which is the ``correct'' answer. 
	
\subsubsection*{Regression Analysis of Personal Factors}
\label{sec:linear-regression}
The number of years working with CVSS did not affect performance, nor did self-rated knowledge of CVSS documents (\Cref{tab:linear-regression}). 
Usage of the online calculator and internal company documents during the assessments has no influence, whereas usage of CVSS documentation led to a better performance. The most important factor is the number of correct answers to the seven cases from CVSS documentation.
There is slight evidence that people who rated themselves as intermediate in their knowledge of CVSS performed better than those who rated themselves as having basic or no knowledge. However, this effect is not significant. The attendance of FIRST training courses and opinion on CVSS usage have no effect. Respondents that think that the BaseScore should be used without adaption have better results than those who think that adaptions should be used. $R^2$ is 0.118 which means that almost 12\% of variance in the dependent variable can be explained by the explanatory model, which can be interpreted as low explanatory power~\cite{1977-cohen}. 
	
\begin{table}
	\begin{center}
		\begin{tabularx}{0.9\linewidth}{X|c}
			\toprule
			Factor & Std. score\\
			\midrule
			Years working with CVSS & 0.0133\\
			& (0.71)\\
			Self-assessed knowledge of CVSS documents & -0.0907\\
			& (-1.00)\\
			Using onlince caluclator during CVSS assess- & 0.106\\
			ment (Ref. Not selected) & (0.59)\\
			Using CVSS documents during CVSS assess- & 0.423$^{*}$\\
			ment (Ref. Not selected) & (2.42)\\
			Using internal company documents during & -0.0157\\
			CVSS assessment (Ref. Not selected) & (-0.10)\\
			Correct answers to CVSS documentation cases  & 1.163$^{**}$\\
			 & (3.06)\\
			Self-assessed expertise in CVSSv3.1 &\\
			(Ref. Low expertise)&\\
			\hspace*{2em}Intermediate & 0.366$^{+}$\\
			& (1.76)\\
			\hspace*{2em}Expert & 0.349\\
			& (1.54)\\
			Attended FIRST training course & -0.260\\
			(Ref. Yes) & (-0.84)\\
			Opinion: CVSS should be used to assess sever- & 0.0699\\
			ity (Ref. Not selected) & (0.49)\\
			Opinion: Using Base Score with adaptation & 0.369$^{*}$\\
			(Ref. Without adaptation) & (2.31)\\
			Constant & -0.973$^{+}$\\
			& (-1.97)\\
			\midrule
			Observations & 196\\
			$R^2$ & 0.118\\
			Adjusted $R^2$ & 0.065\\
			\bottomrule
		\end{tabularx}
		\end{center}
	\caption{Linear regression of personal factors with standardized score as dependent variable (\Cref{sec:quant-analysis})}
	\label{tab:linear-regression}
\end{table}
	
\subsection{Attitudes towards CVSS (RQ4)}

\begin{figure*}
	\begin{center}
		\includegraphics[width=0.8\textwidth]{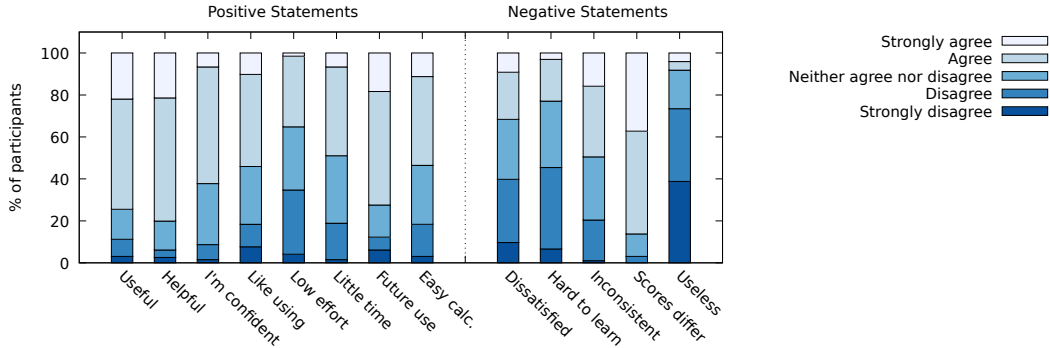}
	\end{center}
	\caption{Ratings of usability statements (N = 196); full statements are presented in \Cref{tab:results-usability} in Appendix~\ref{sec:usability-statements}}
	\label{fig:results-usability}
\end{figure*}

Attitudes towards CVSS signaled that it is both problematic and useful. This split can be discerned in results of the questionnaire and open feedback comments. As shown in \Cref{fig:results-usability}, three quarters agreed that CVSS is useful, while almost two thirds felt confident using it. The assessment of whether it required low effort was evenly split. Furthermore, almost half of the respondents considered scores to be inconsistent and an overwhelming majority agreed that scores differ across different raters. These ratios suggest that users are aware of problems associated with CVSS, but that these do not outweigh its advantages. Over 70\% clearly agreed to use CVSS in the future.

Several respondents used free-text comments to relay perceived advantages as well as discontent with CVSS. Comments ranged from proclaiming CVSS ``misleading'' (P19)
and offering merely a \emph{``veneer of objectivity used to justify whatever subjective opinion the scorer brings with them''} (P124)
to thoughtful appreciation of its capabilities and shortcomings. Respondents problematized the subjective basis for evaluations, the pitfalls of generalization, and the lack of applicability regarding particular kinds of evaluation targets, such as hardware or medical devices. 

CVSS  acceptance and recognition in the industry was considered a boon, but it could also lead to CVSS being misused: \emph{``for certain vulnerability types, CVSS is a good scoring system. However, there are so many aspects of security that CVSS is not suitable for scoring, yet clients still request CVSS to be used as it's ``known'', even where it is not suitable''} (P152).
Some respondents saw advantages of CVSS especially in the ability to clearly communicate severity to customers, its usefulness when coupled with other forms of evaluation, and in its ability to serve as a tool of ``rough guidance'' (P186)
for vulnerability prioritization. 
One commenter clarified, \emph{``While I like CVSS for the questions it asks, I dislike the scoring''} (P179).

Another problem mentioned by the participants is the NVD and the CVE reporting. Seen as main reference, the NVD does not seem to follow the CVSS guidelines strictly, which often leads to confusion, according to the participants. Also, the provided vulnerability descriptions are often missing important information. One respondent stated that most of the people who provide the score for the CVE reporting lack understanding of some CVSS aspects and went on to add, \emph{``This bad input data which is perpetuated by bad CVE reporting is, to my mind, a far bigger risk to the quality, reputation, and trustworthiness of CVSS''} (P37).

The split assessment of CVSS as both problematic and useful is encapsulated in this insightful comment, which highlights it as the best available measure while clearly gauging its shortcomings:
\emph{``Internally we often have discussions about each issue and sometimes disagree on the outcome. It's usually an exercise of applying a score and consulting the internal and external guides for ``interpretation'' of the issue. [$\dots$] While it's not perfect, I think CVSS is the best measure we have as an industry for communicating severity of issues.''} (P51).

\subsection{Consistency of Evaluations over Time (RQ5)}
\label{sec:follow-up}

In the follow-up survey, we investigated if a participant chose the same CVSS evaluation for the same vulnerability as in the main survey (RQ5). 
Each participant evaluated four vulnerabilities: two that they evaluated in the main survey, and two that were evaluated in the main survey by other group but were new to them. The unfamiliar vulnerabilities were added as distractors.
We included Banner Disclosure (evaluated by Group 1 in the main survey) and HTTPOnly (evaluated by Group 2) into the follow-up survey to  investigate if security deficiencies are evaluated consistently over time (see \Cref{sec:discussion-session}). For Group 1 we chose Stored XSS as it was evaluated less consistently than Reflected XSS in the main survey. We selected Adobe Acrobat for Group 2 as it was evaluated less consistently than Google Chrome. 

The follow-up survey was conducted in fall 2021, nine months after the main survey. We contacted all participants who agreed to be contacted for this purpose (115 out of 196), and received 59 responses: 32 from Group 1 and 27 from Group~2. 
\Cref{tab:results-followup} shows the comparison of the evaluations by the same participant: For Group 1, approximately 35\% (Banner Disclosure) to 40\% (Stored XSS) of the participants scored a different severity. For Group 2, 45\% (Adobe Acrobat) to 55\% (HTTPOnly) of the participants evaluated the severity different. 
This noteworthy amount of inconsistency indicates that evaluations are not consistent over time.

\begin{table}[t]
	\begin{tabularx}{\linewidth}{l|cc c}
		\toprule
		& Same & Lower & Higher\\
		Vulnerability & Severity & Severity & Severity\\
		\midrule
		StorXSS (N = 32) & 19 & 5 & 8\\
		Banner Disclosure (N = 32) & 21 & 5 & 6\\
		Adobe Acrobat (N = 27) & 15 & 6 & 6\\
		HTTPOnly (N = 27) & 12 & 8 & 7\\
		\bottomrule
	\end{tabularx}
	\vspace*{0.5em}
	\caption{Comparison of evaluation severities between the main survey and the follow-up survey.}
	\label{tab:results-followup}
\end{table}

\section{Discussion}
\label{sec:discussion}
We found that evaluations are not consistent across different evaluators, and are also not consistent over time for the same evaluator. We found a low number of personal factors related to individual (in)consistency, and the corresponding regression model had a low explanatory power.
Especially years of experience, found by Allodi et al.~\cite{2020-allodi} to be related to the accuracy of evaluations, did not have an effect in our study. 
It seems that inconsistency is more closely related to the properties of CVSS, such as problematic metrics and documentation, than to the personal factors that we investigated. In the following, we present the main areas in which we recommend CVSS improvements.

\subsubsection*{Steps towards Effective Vulnerability Management}
Our findings have important implications for vulnerability management. We highlighted that a significant number of participants incorporate CVSS in their risk management processes and vulnerability prioritisation methods. Consequently, these inconsistencies in CVSS scores can lead to inaccurate resource allocation. This could result in critical vulnerabilities being sidelined while less severe ones receive undue attention. This discrepancy poses a serious challenge to effective vulnerability management, potentially introducing risks that could escalate to tangible damages.

Our research also covered security deficiencies, weaknesses that are not vulnerabilities in the traditional sense because they cannot be directly exploited. Despite their indirect nature, these issues are often assessed using CVSS, a practice that our study shows to be flawed. This necessitates that FIRST either elucidates clear guidelines on how to handle these security deficiencies within the CVSS framework or provides an explicit directive that such issues are not suitable for CVSS assessment. In fact, the current framework may lead to an over-prioritisation of these second-level mitigations, potentially distracting from urgent vulnerabilities. Given this widespread practice, there is an urgent need for action to address this discrepancy.

\subsubsection*{Enhancing Accessibility of CVSS Documentation}

Our study also reveals that a significant proportion of evaluators rarely consult the CVSS documentation, with around 30\% of our sample reporting that they have never read it. Notably, the accuracy of CVSS evaluations did not show a direct correlation with self-reported familiarity with the documentation. However, evaluators who reported consulting the documentation during their evaluations showed better  performance, suggesting that a solid understanding of the documentation can  improve the accuracy of CVSS ratings.

As we have not conducted an empirical study of the reasons why users neglect the CVSS documentation, we propose several hypotheses. One possible barrier could be the dispersion of information across three documents: the specification, the user guide and the examples. This may discourage users from  exploring the documentation, as they may perceive extraction and synthesis of the essential content as too laborious.

Moreover, the widespread use of FIRST's online CVSS calculator by 77\% of the participants offers another plausible explanation. The calculator provides instant access to tooltips that show brief explanations for each metric, which may give users a sense of being adequately informed. This  might lead users to believe that delving into the documentation would not provide any further valuable insights. We strongly recommend empirical investigation of these proposed factors as part of future work to gain a more accurate understanding of the issue.

In light of these findings, we recommend making the CVSS documentation more accessible and intuitive to use. The online calculator provides a unique opportunity for this improvement. This tool, which currently provides excerpts from the specification and brief explanations, could be further optimised to provide evaluators with more comprehensive and contextually relevant guidance. This opens a promising avenue for future research.

In addition, given that guidance on difficult rating cases is currently spread across several documents, centralising this information would eliminate the need for time-consuming cross-referencing, thereby making the evaluation process more efficient and helping evaluators to apply CVSS more consistently and accurately.

\subsubsection*{Refining Metrics}

Our study results also indicate that specific metrics such as User Interaction, Attack Vector and Scope require more precise definition and clearer guidance for some widespread vulnerabilities.  We suspect that similar ambiguities may exist for other types of vulnerabilities and other metrics, highlighting the need for further research.

In particular, the Scope metric presents a significant challenge. We found that irrespective of the actual vulnerability type, a Changed Scope (S:C) was applied by about one third of participants, while the remaining two thirds consistently opted for an Unchanged Scope (S:U). The recurring selection of an Unchanged Scope (S:U) by a considerable number of participants appears to stem from their challenges in interpreting the Scope metric, as evidenced by their comments during our study. This strongly suggests that the Scope metric requires a comprehensive reassessment to improve its understandability and ensure its consistent application across varying vulnerabilities.

\subsubsection*{Balancing the Value and Shortcomings of CVSS}

Despite the challenges identified with CVSS, our findings confirm its value as an indispensable tool within the security community. Participants largely valued CVSS for its role as a standardised mechanism for communicating the severity of vulnerabilities to different stakeholders. They recognised its limitations, but continued to use it, highlighting the lack of better alternatives. To capture this sentiment, we cite P124: \emph{``CVSS is like democracy: the worst system available, except for all the other systems ever tried.''}. As we move forward, it is clear that while CVSS has its problems, the focus should be on refining and improving this system, rather than discarding it altogether. This highlights the critical need for continued research and iterative development to improve the accuracy, consistency and usability of CVSS, ultimately supporting more robust vulnerability management practices.
\subsubsection*{Limitations}
\label{sec:limitations}

This study, although carefully designed, has several limitations. Thus, CVSS discussion experts and survey experts contributed their subjective perceptions of CVSS. With different experts, the focus and vulnerabilities of our study might have been different. To mitigate this limitation, we asked for each vulnerability if the participants rate similar types of vulnerabilities in their daily work. As most participants agreed (\Cref{tab:matrix-questions}), we assume that selected vulnerabilities are fairly common. Further, five of the selected vulnerabilities are widespread and particularly critical according to the 2022 CWE Top 25 list.

Vulnerability assessment process was limited compared to real life, as evaluators usually possess in-depth knowledge about the vulnerability. This might have influenced the consistency of evaluations.
As we only considered participants who are currently using CVSS, all persons were excluded who do not currently use CVSS -- even those who are familiar with CVSS but decided not to use it. Thus, the participants might not represent the whole community of vulnerability assessors.
Demographics of the participants had a low diversity, as most participants identified
as male and currently live in Germany or the USA.
However, it is unclear which demographic distribution should be targeted, as there are no demographic statistics about regular CVSS users. Possibly we did not find all significant effects due to limitations regarding the sample size of 196 participants.
\section{Conclusion}
\label{sec:conclusion}

We investigated the consistency of CVSSv3.1 evaluations focusing on specific metrics for widespread vulnerabilities. None of the considered cases were evaluated consistently.
However, we found only slight evidence that personal factors influence the consistency of evaluations.
The follow-up study showed that evaluations of one evaluator are not consistent over time. Still, most participants have a positive attitude of CVSS, despite their awareness of its flaws. 

In the future, a similar study design could be used to investigate CVSS consistency for different vulnerability types, e.g., in embedded systems. Additional influencing factors, such as risk propensity, intuition or emotional state, could be investigated. The research on the consistency of metrics could also be expanded to the metrics which were not covered by this study. The Environmental and Temporal Scores could also be investigated.

\subsubsection*{Public Data Set} Together with this paper, we release a public data set with questionnaires, descriptive results and pseudonymized datasets from the main study and follow-up study \cite{public-data-set}.

\subsubsection*{Acknowledgments}
We thank FIRST for permission to use the code of the CVSSv3.1 online calculator and Andreas Hammer for embedding this code into LimeSurvey. We thank the anonymous reviewers whose thorough comments greatly improved this paper, as well as the anonymous shepherd for guidance. We thank Thomas Schreck for substantial support with selection of vulnerabilities and defining the ground truth, and Jonathan M. Spring for suggesting important improvements of survey design. We thank Kilem L. Gwet for suggesting Finn's coefficient as the most appropriate agreement indicator and Yury Zablotski for calculating it. This work was partially funded by the German Federal Ministry of Education and Research under grant 16KIS1271K.

\bibliographystyle{IEEEtran}
\bibliography{cvss_paper}

\appendices

\section{Clarity of Description and Rating of Similar Vulnerabilities}
\label{sec:appendix-table-matrix-questions}

\Cref{tab:matrix-questions} presents participants' ratings of clarity of the vulnerability descriptions and whether they often rate similar vulnerabilities in their daily work.

\begin{table}[h]
	\begin{center}
		\begin{tabularx}{\linewidth}{l|c c c|c c c}
			\toprule
			& \multicolumn{3}{c|}{Description understandable} & \multicolumn{3}{c}{Rating similar vulns.}\\
			& $\mu$ & $\sigma$ & $M$ & $\mu$ & $\sigma$ & $M$\\
			\midrule
			Stored XSS & 3.23 & 1.17 & 4 & 3.48 & 1.17 & 4\\
			Reflected XSS & 3.33 & 1.16 & 4 & 3.48 & 1.19 & 4\\
			SQL Injection & 3.93 & 0.86 & 4 & 3.73 & 1.06 & 4\\
			Banner Disclosure & 3.77 & 1.05 & 4 & 3.60 & 1.10 & 4\\
			Adobe Acrobat & 4.09 & 0.77 & 4 & 3.41 & 1.13 & 4\\
			Google Chrome & 3.92 & 0.82 & 4 & 3.40 & 1.07 & 4\\
			MITM MyPalette & 3.81 & 0.85 & 4 & 3.44 & 1.09 & 4\\
			HTTPOnly & 3.75 & 0.93 & 4 & 2.97 & 1.11 & 4\\
			\bottomrule
		\end{tabularx}
		\caption{Participants indicated on a 5-point Likert scale (1 = ``Strongly disagree'', $\dots$, 5 = ``Strongly agree'') their agreement or disagreement with the statements ``The description of the security issue is clear and understandable'' and `` I often evaluate security issues of this type in my
daily work''. $\mu$ = mean, $\sigma$ = standard deviation, $M$ = median.}
		\label{tab:matrix-questions}
	\end{center}
\end{table}

\section{Vulnerability Descriptions}
\label{sec:appendix-questionnaire-vuln-descriptions}
The descriptions of the vulnerabilities of the main survey are depicted in \Cref{tab:vuln-descriptions}. Adjustments within the descriptions are written in italics.

\begin{table*}[ht]
	\begin{tabularx}{\linewidth}{l X}
		\toprule
		Vulnerability & Description, vector string and CVE\\
		\midrule
		\multirow{1}{0.1\linewidth}{Reflected XSS} & Open edX \emph{in version} Ironwood.1 \emph{is vulnerable to a reflected XSS attack. An unauthenticated attacker is able to manipulate the HTTP URI parameter} /support/certificates?course\_id=.\\
		& NVD vector string: CVSS:3.1/AV:N/AC:L/PR:N/UI:R/S:C/C:L/I:L/A:N, Score: 6.1, CVE-2019-20512\cite{cve-2019-20512}\\
		\midrule
		\multirow{1}{0.1\linewidth}{Stored XSS} & Studio in Open edX Ironwood 2.5 allows users to upload SVG files via the ``Content>File Uploads'' screen. These files can contain JavaScript code and thus lead to Stored XSS.\\
		& NVD vector string: CVSS:3.1/AV:N/AC:L/PR:L/UI:R/S:C/C:L/I:L/A:N, Score: 5.4, CVE-2020-13145~\cite{cve-2020-13145}\\
		\midrule
		\multirow{1}{0.1\linewidth}{SQL Injection} & A vulnerability in the web-based management interface of Cisco Prime Collaboration Provisioning Software allows an \emph{authenticated attacker} to conduct \emph{SQL injection attacks}. The vulnerability exists because the web-based management interface improperly validates user input for specific SQL queries. An attacker \emph{can} exploit this vulnerability by authenticating to the application with valid administrative credentials and sending malicious requests to an affected system. A successful exploit \emph{allows} the attacker to view information that they are not authorized to view, make changes to the system that they are not authorized to make, or delete information from the database that they are not authorized to delete.\\
		& NVD vector string: CVSS:3.1/AV:N/AC:L/PR:H/UI:N/S:U/C:H/I:H/A:H, Score: 7.2, CVE-2020-3184~\cite{cve-2020-3184}\\
		\midrule
		\multirow{1}{0.1\linewidth}{Banner Disclosure} & A vulnerability in the web-based management interface of Cisco Prime Collaboration Provisioning \emph{allows} an unauthenticated attacker to obtain sensitive information about an affected device. The vulnerability exists because replies from the web-based management interface include unnecessary server information. An attacker could exploit this vulnerability by inspecting replies received from the web-based management interface. A successful exploit \emph{allows} the attacker to obtain details about the operating system, including the web server version that is running on the device, which could be used to perform further attacks.\\
		& NVD vector string: CVSS:3.1/AV:N/AC:L/PR:N/UI:N/S:U/C:L/I:N/A:N, Score: 5.3, CVE-2020-3193~\cite{cve-2020-3193}\\
		\midrule
		\multirow{3}{0.1\linewidth}{Adobe Acrobat Buffer Overflow} & Adobe Acrobat and Reader version 9.0 and earlier are vulnerable to a buffer overflow, caused by improper bounds checking when parsing a malformed JBIG2 image stream embedded within a crafted PDF document. \emph{The attacker can overflow a buffer and execute arbitrary code on the system with the privileges of the user or cause the application to crash.}\\
		& Examples vector string: CVSS:3.1/AV:L/AC:L/PR:N/UI:R/S:U/C:H/I:H/A:H, Score: 7.8, CVE-2009-0658~\cite[p. 26]{cvssexamplev3.1}\\
		\midrule
		\multirow{3}{0.1\linewidth}{Google Chrome PDFium JPEG} & This vulnerability allows \emph{attackers} to execute arbitrary code on vulnerable installations \emph{of Google Chrome}. The specific flaw exists within the handling of JPEG 2000 images. A specially crafted JPEG 2000 image embedded inside a PDF cain preliminary surveyn force Google Chrome to write memory past the end of an allocated object. An attacker can leverage this vulnerability to execute arbitrary code under the context of the current process.\\
		& Examples vector string: CVSS:3.1/AV:N/AC:L/PR:N/UI:R/S:U/C:H/I:H/A:H, Score: 8.8, CVE-2016-1645~\cite[p. 36]{cvssexamplev3.1}\\
		\midrule
		\multirow{1}{0.1\linewidth}{MITM MyPalette} & Android App ’MyPallete’ and some of the Android banking applications based on ’MyPallete’ do not verify X.509 certificates \emph{from servers, which allows} man-in-the-middle attackers to spoof servers and obtain sensitive information via a crafted certificate.\\
		& NVD vector string: CVSS:3.1/AV:N/AC:H/PR:N/UI:N/S:U/C:H/I:H/A:N, Score: 7.4, CVE-2020-5523~\cite{cve-2020-5523}\\
		\midrule
		\multirow{1}{0.1\linewidth}{HTTPOnly} & Synology Router Manager (SRM) before 1.2.4-8081 does not include the HTTPOnly flag in a Set-Cookie header for the session cookie, which makes it easier \emph{for attackers} to obtain potentially sensitive information via script access to this cookie.\\
		& NVD vector string: CVSS:3.1/AV:N/AC:L/PR:N/UI:R/S:C/C:L/I:L/A:N, Score: 6.1, CVE-2020-27658~\cite{cve-2020-27658}\\
		\bottomrule
	\end{tabularx}
	\caption{Vulnerability descriptions of the main survey. Adjustments within the descriptions are written in italics.
	}
	\label{tab:vuln-descriptions}
\end{table*} 

\section{Usability Statements}
\label{sec:usability-statements}

Usability statements for \Cref{fig:results-usability} are presented in \Cref{tab:results-usability}.

\begin{table}[ht]
	\begin{center}
		\begin{tabularx}{\linewidth}{l|l}
			\toprule
			Label & Statement\\
			\midrule
			\texttt{Useful} & CVSS is useful for vulnerability management.\\
			\texttt{Helpful} & CVSS is helpful for assessing vulnerabilities.\\
			\texttt{I'm confident} & I feel confident using CVSS.\\
			\texttt{Like using} & I like using CVSS.\\
			\texttt{Low effort} & CVSS evaluations require low effort.\\
			\texttt{Little time}& CVSS evaluations take little time.\\
			\texttt{Future use} & Personally, I would like to use CVSS in the future.\\
			\texttt{Easy calc.} & CVSS scores are easy to calculate after\\
			& understanding the vulnerabilities.\\
			\texttt{Dissatisfied} & I feel dissatisfied with CVSS.\\
			\texttt{Hard to learn} & CVSS is hard to learn.\\
			\texttt{Inconsistency} & There is too much inconsistency in CVSS.\\
			\texttt{Scores differ} & CVSS scores differ depending on who calculates.\\
			\texttt{Useless} & CVSS scores are useless.\\
			\bottomrule
		\end{tabularx}
		\caption{Rated usability statements for \Cref{fig:results-usability}}
		\label{tab:results-usability}
	\end{center}
\end{table}
\section{Questionnaire of the Main Survey}
\label{sec:appendix-questionnaire-mainstudy}
Questionnaire starts with a message informing the participants about requirements for participation (currently assessing vulnerabilities with CVSS), the purpose of the study, followed by the participation consent.

\subsection{Preselection Question}
Are you currently assessing vulnerabilities using CVSS? (Yes/No)
(If ''No`` is selected, the survey ends with a short explanation that we seek participants who currently assess vulnerabilities using CVSS.)

\subsection{Your work experience}
\begin{itemize}
	\item What is your current job title or position: (free text)
	\item Please estimate how many years you have been working in your current position.
	\item Please assess your expertise in the following areas:\\
	(None, Basic, Intermediate, Advanced, Expert)
	\begin{itemize}
		\item System Software Security / Desktop Application Software Security / Mobile Application Software Security
		/ Web Application Software Security / Database Security / Embedded Systems Security / Network Security / Product Security
	\end{itemize}
	\item Please estimate how many years you have been working in the IT security sector.
	\item Please estimate how many years you have been working with CVSS.
	\item Please assess your knowledge of CVSS:\\
	(None, Basic, Intermediate, Advanced, Expert)
	\begin{itemize}
		\item CVSSv3.1 / CVSSv3.0 / CVSSv2
	\end{itemize}
	\item FIRST offers special courses to get familiar with CVSS. Have you participated in the following FIRST e-learning CVSS courses? (Yes/No)
	\begin{itemize}
		\item FIRST Mastering CVSS v3.1 / v3.0
	\end{itemize}
\end{itemize}

\subsection{CVSS at your work}
\begin{itemize}
	\item What are your default CVSS versions for daily tasks? (multiple-choice)
	\begin{itemize}
		\item CVSSv3.1 / CVSSv3.0 / CVSSv2 
		\item I don't have a default CVSS version.
		\item I don't use CVSS for my daily tasks.
	\end{itemize}
	\item For whom do you assess vulnerabilities using CVSS? (multiple-choice)
	\begin{itemize}
		\item For customers (e.g., for other companies)
		\item Internally (e.g., for your own company, for yourself)
		\item Other, please indicate: (free text)
		\item I don’t assess vulnerabilites.
	\end{itemize}
	\item How is CVSS used at your work? (multiple-choice)
	\begin{itemize}
		\item Prioritising vulnerabilities
		\item Assessing the severity of vulnerabilities
		\item Assessing the risk of vulnerabilities
		\item Other, please indicate: (free text)
		\item I don’t know.
		\item We don’t use CVSS
	\end{itemize}
	\item How is the CVSS Base Score used at your work primarily?
	\begin{itemize}
		\item Without adaptation / adapted by the Environmental Score / adapted by the Temporal Score / adapted by the Environmental and Temporal Score.
		\item Other, please indicate: (free text)
		\item I don’t know.
		\item We don’t use the Base Score.
	\end{itemize}
	\item What documents or tools are you using during a CVSS assessment? (multiple-choice)
	\begin{itemize}
		\item Online-Calculator (e.g., FIRST’s CVSS-Calculator)
		\item CVSS Specification / User Guide / Examples 
		\item Internal company specific documents and software
		\item Other, please indicate: (free text)
		\item Nothing
	\end{itemize}
	\item Please estimate your average time (in minutes) spent per security issue while evaluating using CVSS.
	\item Please estimate the number of security issues you are evaluating using CVSS on average per week.
	\item How many persons (including yourself) are usually involved in an evaluation using CVSS at your work?
	\item If you are unsure about a CVSS assessment, what or whom do you consult? (multiple-choice)
	\begin{itemize}
		\item CVSS Specification / User Guide / Examples
		\item Internet research (e.g., Google, Stack Overflow)
		\item This is an attention test. Please check this item additionally to any other items that you checked here.
		\item Ratings of the same issue by other parties
		\item Ratings of similar issues
		\item Ask coworkers
		\item Other, please indicate: (free text)
		\item I don't consult anybody or anything.
	\end{itemize}
	\item Are Base Scores that you calculated reviewed or verified by coworkers or other people?
	\begin{itemize}
		\item Always / Often / Occasionally / Rarely / Never / Don't know
	\end{itemize}
\end{itemize}

\subsection{FIRST's official documents of CVSS}
\begin{itemize}
	\item Please assess your knowledge of FIRST’s official documents of CVSS:\\
	(None, Basic, Intermediate, Advanced, Expert)
	\begin{itemize}
		\item Specification CVSSv3.1 / CVSSv3.0 / CVSSv2
		\item User Guide CVSSv3.1 / CVSSv3.0 / CVSSv2
		\item Examples CVSSv3.1 / CVSSv3.0 / CVSSv2
	\end{itemize}
	\item Please indicate when you have last consulted FIRST’s official documents:\\
	(Never, More than a year ago, A few months ago, A few weeks ago, A few days ago or today)
	\begin{itemize}
		\item Specification CVSSv3.1 / CVSSv3.0 / CVSSv2
		\item User Guide CVSSv3.1 / CVSSv3.0 / CVSSv2
		\item Examples CVSSv3.1 / CVSSv3.0 / CVSSv2
	\end{itemize}
\end{itemize}

\subsection{CVSS scoring assessment}
(This part of the survey asked the participants to evaluate four vulnerabilities with randomized order using CVSSv3.1. The same questions were asked for each vulnerability.)

\begin{itemize}
	\item Please use this embedded version of FIRST's Online Calculator to compute the Base Score.
	\begin{itemize}
		\item[] \includegraphics[width=\linewidth]{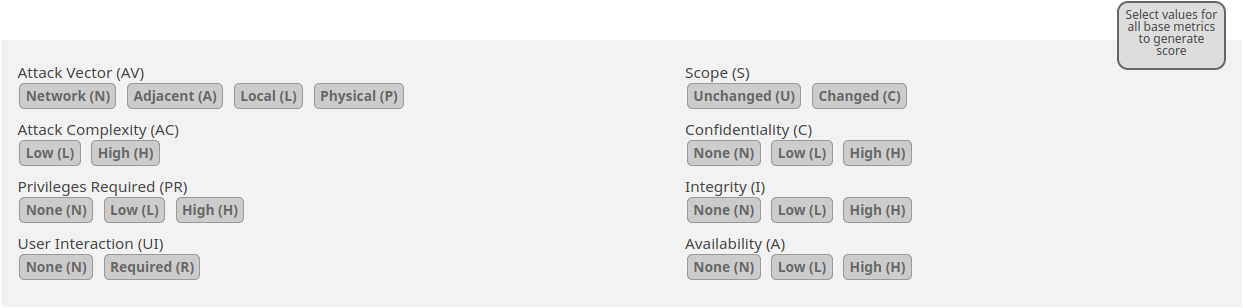}
	\end{itemize}
	\item If you relied on any special assumptions in your assessment please state them here: (free text)
	\item Please indicate your agreement with the following statements:\\
	(Strongly disagree, Disagree, Neither agree nor disagree, Agree, Strongly agree)
	\begin{itemize}
		\item I'm fully confident in my calculation of the Base Score.
		\item I know what the software described in the security issue does.
		\item The description of the security issue is clear and understandable.
		\item The described security issue is fully suitable for an evaluation using CVSS.
		\item I often evaluate security issues of this type in my daily work.
		\item This is an attention test. Please click ''Agree`` here. (This question only appeared for one vulnerability)
	\end{itemize}
	\item Please leave a comment in case there was an ambiguity or any other issues with the description or the calculation of the security issue. If you think CVSS is unsuitable to evaluate this security issue, please explain why: (free text)
\end{itemize}

\subsection{CVSS documentation questions}
\label{sec:unklarheitenFragen}
\begin{itemize}
	\item In case of a CVSS assessment I usually evaluate the impact metrics considering$\dots$
	\begin{itemize}
		\item $\dots$ a realistic scenario (e.g., software or company specific characteristics).
		\item $\dots$ the reasonable worst case scenario. (correct\footnote{Correct answers according to CVSS documentation~\cite{cvssuserv3.1,cvssspecsv3.1,cvssexamplev3.1}.})
		\item Other, please indicate: (free text)
		\item I don’t understand this question.
	\end{itemize} 
	\item If an attacker requires advanced knowledge to exploit the vulnerability, I usually evaluate the vulnerability assuming$\dots$
	\begin{itemize}
		\item $\dots$ the attacker already has advanced knowledge. (correct)
		\item $\dots$ the attacker does not yet have the advanced knowledge.
		\item Other, please indicate: (free text)
		\item I don’t understand this question. 
	\end{itemize}
	\item 	If a system needs to be in a specific configuration in order to exploit the vulnerability, I usually evaluate assuming$\dots$
	\begin{itemize}
		\item $\dots$ the system is not in this specific configuration and the attacker accordingly needs to prepare the system.
		\item $\dots$ the system is already in this specific configuration. (correct)
		\item Other, please indicate: (free text)
		\item I don’t understand this question. 
	\end{itemize}
	\item If Social Engineering is required in order to exploit the vulnerability, I usually adapt$\dots$
	\begin{itemize}
		\item $\dots$the User Interaction metric. (correct)
		\item $\dots$the Privileges Required metric.
		\item $\dots$none.
		\item Other, please indicate: (free text)
		\item I don’t understand this question.
	\end{itemize}
	\item If an attacker can cause moderate damage with unprivileged user rights or can cause serious damage with admin rights, I usually evalutate the Privileges Required metric as$\dots$
	\begin{itemize}
		\item Privileges Required: None.
		\item Privileges Required: Low.
		\item Privileges Required: High. (correct)
		\item Other, please indicate: (free text)
		\item I don’t understand this question.
	\end{itemize} 
	\item If a scope change (S:C) occurred, I usually evaluate the impact on$\dots$
	\begin{itemize}
		\item $\dots$ the vulnerable component.
		\item $\dots$ the impacted component.
		\item $\dots$ the component which suffers the most severe outcome. (correct)
		\item Other, please indicate: (free text)
		\item I don’t understand this question. 	
	\end{itemize}
	\item The metric Availability reflects the availability of$\dots$
	\begin{itemize}
		\item $\dots$ performance and operation of the impacted component. (correct)
		\item $\dots$ data used by the impacted component.
		\item $\dots$ data, performance and operation of the impacted component.
		\item Other, please indicate: (free text)
		\item I don’t understand this question. 
	\end{itemize}
\end{itemize}

\subsection{Your opinion about CVSS}
\begin{itemize}
	\item In your opinion, how should CVSS be used? (multiple-choice)
	\begin{itemize}
		\item Assessing the severity of vulnerabilities / Assessing the risk of vulnerabilities / Prioritising vulnerabilities
		\item Other, please indicate: (free text)
	\end{itemize}
	\item In your opinion, how should the Base Score be used primarily?
	\begin{itemize}
		\item Without adaptation / adapted by the Environmental Score / adapted by the Temporal Score / adapted by the Environmental and Temporal Score
		\item Other, please indicate: (free text)
	\end{itemize}
	\item Please indicate your agreement with the following statements: (statements from \Cref{tab:results-usability}, randomized) 
\end{itemize}

\subsection{Demographic questions}
\begin{itemize}
	\item Please indicate your gender.
	\begin{itemize}
		\item Male / Female / Diverse / Prefer not to say
	\end{itemize}
	\item Please indicate your year of birth. 
	\item Please indicate your country of residence.
	\item Please indicate your current main occupation.
	\begin{itemize}
		\item Employee, civil servant / Self-employed (with employees) / Freelancer / Student / Other, please indicate: (free text) / Prefer not to say
	\end{itemize}
	\item Please indicate your highest school education level.
	\begin{itemize}
		\item No school certificate / Primary school/elementary school or equivalent / Middle school/secondary school or equivalent (not meeting university entrance requirements) / High school or equivalent (meeting university entrance requirements) / Other, please indicate: (free text) / Prefer not to say
	\end{itemize}
	\item Please indicate your highest completed academic/professional education level.
	\begin{itemize}
		\item No completed academic/professional education / Completed vocational training / Bachelor's degree or equivalent / Master's degree or professional degree (M.D., J.D., etc.) or equivalent / Ph.D. (doctoral degree) / Other, please indicate: (free text) / Prefer not to say
	\end{itemize}
	\item Please estimate how many people are employed at your work. 
	\begin{itemize}
		\item 1-9 / 10-49 / 50-249 / 250-499 / 500-999 / 1,000-4,999 / 5,000-9,999 / 10,000 or more / I don't know. / Prefer not to say
	\end{itemize}
	\item Please indicate the economic sector of your company or organization.
	\begin{itemize}
		\item Accommodation and food service activities (e.g., hotel, camping grounds, restaurants, event catering) / Administration and support service activities (e.g., rental leasing, employment activities, office administrations) / Agriculture, forestry and fishing / Arts, entertainment and recreation / Construction (e.g., construction of buildings, civil engineering, demolition) / Education / Electricity, gas, steam and air-conditioning supply / Financial and insurance activities / Human health services, residential care and social work activities / Information and communication (e.g., publishing activities, programming, information service activities) / Manufacturing (e.g., manufacturing of food, textiles, chemical products, electronics, machinery) / Mining and quarrying / Professional, scientific, technical activities (e.g., legal, accounting, research \& development, technical testing and analysis) / Public administration and defence, compulsory social security / Real estate activities (e.g., buying, selling and renting real estates) / Transportation and storage (e.g., passenger transport, warehousing, postal and courier activities) / Water supply, sewerage, waste management and remediation / Wholesale and retail trade, repair of motor vehicles and motorcycles / I don't know. / Other, please indicate: (free text) / Prefer not to say
	\end{itemize}
	\item If possible, please provide the name of your company. This information will not be published and will be only used internally to correlate answers of people from the same company.
\end{itemize}

\subsection{Feedback}
\begin{itemize}
	\item Please give us feedback on the completion of this survey:
	(Strongly disagree, Disagree, Agree, Strongly agree)
	\begin{itemize}
		\item I was distracted during the questioning (e.g., phone calls, other people).
		\item I have answered the questions carefully.
		\item I consulted FIRST's official CVSS documents during this survey.
		\item I tried to get additional information about the security issues (e.g., used Google, asked other people).
	\end{itemize}
	\item If you would like to give us additional feedback on the survey, please share it here.
	\item If you have any thoughts on CVSS which were not covered by this survey, please share them in the text box below. 
	\item Would you like to be informed about the results of this survey? -- Yes / No
	\item May we contact you for a short follow-up CVSS survey? -- Yes / No
	\item If you answered ``Yes'' to at least one of the questions above, please provide a contact email address here. The provided email address will be used solely for the purposes you indicated above and for nothing else. It will be deleted after the completion of this project.
\end{itemize}

\newpage

\section{Meta-Review}
\subsection{Summary}
This paper reports a study of the reliability of CVSS v3.1, including what factors affect the reliability of CVSS scores. A survey of 196 individuals reveals which components of CVSSv3.1 are not reliable and which factors are correlated with inconsistent scores. The paper also finds that evaluators are aware of these issues but view CVSS as a useful metric as it supports some general comparisons and allows easier communication about vulnerabilities.

\subsection{Scientific Contributions}
\begin{itemize}
	\item Provides a New Data Set For Public Use
	\item Addresses a Long-Known Issue
	\item Provides a Valuable Step Forward in an Established Field
\end{itemize}

\subsection{Reasons for Acceptance}
\begin{enumerate}
	\item The paper offers valuable insights into the reliability and consistency issues of the CVSS scoring system. The study uncovers specific CVSS variables that exhibit inconsistency and sheds light on factors contributing to these inconsistencies.
	\item The research not only identifies inconsistencies in CVSS scoring but also provides a foundation for potential improvements. By uncovering the factors contributing to scoring discrepancies, the study offers insights that can guide the development of enhanced scoring methodologies and better support decision-making in vulnerability management.
\end{enumerate}

\subsection{Noteworthy Concerns} 
\begin{enumerate} 
	\item One reviewer raised concerns about survey fatigue among participants and its potential impact on responses, particularly in the study's main design based on the survey results. It is important to address this concern by providing evidence that survey fatigue did not influence the underlying ground truth evaluations.
	\item Reviewers agreed that the paper while providing valuable insights, at times appears superficial due to a focus on methodology and multiple research questions. Further exploration of the findings and their implications is recommended to strengthen the paper's impact. Specifically, discussing the potential impacts of inconsistencies on the broader field of software vulnerability management, investigating the reasons behind users not reading CVSS documentation, and providing a clear roadmap for improving CVSS would enhance the practical implications of the research.
\end{enumerate}

\section{Response to the Meta-Review} 
We thank the reviewers for their detailed and helpful comments and address their noteworthy concerns in the following.

The first noteworthy concern refers to the results of the expert survey (Section 3.2), on which the main study was based. Survey fatigue can hardly be completely avoided. We did our best to keep it as small as possible and to mitigate its consequences. For example, in the expert survey, we showed the 10 vulnerabilities in random order and asked control questions at the end, such as ``I have answered the questions carefully''. The results showed no abnormalities. Furthermore, CVSS experts were unusually committed, as they were personally asked by one of us for help and agreed because they found the purpose of the study exciting.

We agree that understanding the broader impact of CVSS scoring inconsistencies on vulnerability management is critical and will highlight the real-world implications of our study. Inaccurate scoring can misdirect resources, as often severity scores guide the remediation process. For example, our study found instances where defense-in-depth measures, generally considered to be secondary layers of protection, were rated as high severity. This misrepresentation can lead to incorrect prioritization, diverting attention and resources away from more pressing vulnerabilities. We expanded our discussion to reflect this.

Considering investigating the reasons behind users not reading CVSS documentation, we cannot do this empirically anymore, as this would require an additional survey. Instead, we discuss hypothetical reasons and suggest an empirical investigation as future work in discussion section.

We acknowledge that providing a clear roadmap would enhance the practical implications of the research. A clear roadmap requires further studies that  empirically evaluate other CVSS metrics that were not highlighted here. We therefore provide in the discussion section a clearly structured guidance to improve CVSS under the aspects we looked at. For example, certain metrics such as Scope should be refined or documentation should be made more accessible.

\end{document}